\documentclass[11pt,a4paper]{article} % use larger type; default would be 10pt
\pdfoutput=1

\usepackage{jcappub}
\usepackage{amsfonts}
\usepackage{wasysym}
\usepackage{amssymb}
\usepackage{graphicx} % support the \includegraphics command and options
\usepackage{booktabs} % for much better looking tables
\usepackage{array} % for better arrays (eg matrices) in maths
\usepackage{paralist}
\usepackage{verbatim}
\usepackage{subfig}
\usepackage{braket}
\usepackage{color}

\newcommand\T{\rule{0pt}{2.6ex}}       % Top strut
\newcommand\B{\rule[-1.2ex]{0pt}{0pt}} % Bottom strut
\newcommand{\perpphoton}{\ket{A_{\perp}}}
\newcommand{\parphoton}{\ket{A_{\parallel}}}
\newcommand{\axion}{\ket{a}}
\newcommand{\dd}{\text{d}}
\newcommand{\beq}{\begin{equation}}
\newcommand{\eeq}{\end{equation}}
\newcommand{\mug}{\ \mu\text{G}}
\newcommand{\kpc}{\text{ kpc}}

\title{A Cosmic ALP Background and the Cluster Soft X-ray Excess in A665, A2199 and A2255}
\author[a]{Andrew J. Powell}
\affiliation[a]{Rudolf Peierls Centre for Theoretical Physics, University of Oxford,\\1 Keble Road, Oxford, OX1 3NP, United Kingdom}
\emailAdd{Andrew.Powell2@physics.ox.ac.uk}

\abstract{It has been proposed that an excess in soft X-ray emission observed from many galaxy clusters can be explained by conversion into photons of axion-like particles (ALPs) in the cluster's magnetic field. Previously it has been shown that conversion of this primordially-generated background of relativistic ALPs---a cosmic ALP background (CAB)---can explain the observed soft X-ray excess in both the centre and the outskirts of the Coma cluster. Here we extend this work to the three clusters A665, A2199 and A2255. We use a stochastic model of the cluster magnetic field to numerically calculate ALP$-$photon conversion probabilities to predict the CAB-generated soft X-ray flux in these clusters. We compare this flux to ROSAT PSPC observations of the three clusters, and use these observations to constrain the CAB parameter space. We find the CAB can reproduce the magnitude of the observed excess in A2199 and A2255 for the same CAB parameters that match the observed soft excess in the Coma cluster. We also find good fit to the morphology of the excesses in these clusters. Simulation of CAB conversion in the cluster A665 is in mild tension with the other clusters due to producing a small but observable excess at large radii where none is observed. This tension is alleviated considering the uncertainty on predicting the count rate in the ROSAT detector, and on the systematics affecting the magnetic field determination. Overall we find good agreement between the CAB parameters for the four clusters studied so far.}


\begin{document}
\maketitle
\section{Introduction}

Observations have revealed a diffuse excess in soft X-ray emission from a large number of galaxy clusters. This work continues the analysis of whether a cosmic ALP background (CAB) can explain such an excess by ALP$-$photon conversion in the cluster's magnetic field \cite{CAB,Angus:2013sua,Kraljic:2014yta}.

The majority of baryons in a galaxy cluster are in the intra-cluster medium (ICM). The ICM is a hot, multi-keV, ionised gas which emits thermally across the X-ray spectrum through thermal bremsstrahlung. In a large number of galaxy clusters excess emission above this expected thermal emission has been seen in the soft X-ray regime ($\sim 0.1-0.4\text{ keV}$). This excess emission was first seen in 1996 from the Coma and Virgo clusters with the Extreme Ultra-Violet Explorer (EUVE) satellite \cite{Lieu1996a,Lieu1996b,Bowyer96}. It has been subsequently observed in many other clusters with EUVE, and the X-ray satellite ROSAT. It has also been seen in a small number of cases with the newer XMM-Newton, Suzaku and Chandra satellites. In a sample of 38 galaxy clusters \cite{astroph0205473} studied using the ROSAT satellite, 13 showed a statistically significant, diffuse, cluster-wide excess, several more clusters showed excesses at low significance. The cluster soft X-ray excess is thus a widespread phenomenon and, interestingly, proposed astrophysical explanations all run into difficulties. For a review of the soft X-ray excess, see \cite{08010977}.

It was proposed in \cite{CAB} that the soft X-ray excess can be explained by ALP$-$photon conversions of a CAB. The excess is produced by a primordially-generated, homogeneous background of relativistic ALPs at $\sim 0.2\text{ keV}$. These ALPs convert to photons of the same energy in the magnetic field of the galaxy cluster. The cluster will then source a large flux of soft X-ray photons from a CAB, in addition to the usual thermal emission from the ICM. The magnitude and morphology of the soft excess in each cluster will be dependent on the magnetic field of the individual cluster, potentially explaining why the excess is not seen in all clusters.

The existence of a CAB is well motivated in string theory models of the early universe \cite{13041804}, where moduli fields drive reheating. The moduli typically decay with a large branching fraction to very weakly coupled, massless ALPs (string axions) which form the CAB \cite{12083562,12083563,13047987,Angus:2013zfa,Angus:2014bia,Hebecker:2014gka}. The decay of the modulus in the expanding universe generates a quasi-thermal energy spectrum for the ALP background. This CAB forms part or all of the dark radiation, which is any hidden species which contributes as radiation energy density. The amount of dark radiation, and thus the energy density of a CAB, is constrained by CMB and BBN observations \cite{2013PASA...30...29R}. Dark radiation is conventionally parameterised in terms of `excess effective number of neutrino species', $\Delta N_{\text{eff}}$. In terms of the dark radiation energy density, it is defined as $\Delta N_{\text{eff}}\equiv\frac{8}{7}\left(\frac{4}{11}\right)^{-4/3}\frac{\rho_{\text{dark}}}{\rho_{\gamma}}$. Current measurements give hints at the $1-2.5\,\sigma$ level for non-zero dark radiation energy density, corresponding to $\Delta N_{\text{eff}}\sim 0.5$ \cite{Ade:2013zuv,Izotov:2014fga}. Future measurements should continue to bring down the uncertainty on $\Delta N_{\text{eff}}$.

The relevant ALP Lagrangian is
\beq
\label{eq:alpcoupling}
	\mathcal{L} \supset \frac{1}{2}\partial_{\mu}a\,\partial^{\mu}a+\frac{1}{4}g_{a\gamma\gamma}\,a\,F\tilde{F},
\eeq
with $g_{a\gamma\gamma}=M^{-1}$ the coupling between photons and ALPs. This coupling allows ALPs and photons to convert into one another in the presence of an external electric or magnetic field \cite{Raffelt}. The ICM of galaxy clusters supports turbulent magnetic fields with $\mathcal{O}(1-10\ \mu G)$ field strength, coherent over a range of scales $\mathcal{O}(1-100\ \text{kpc})$ \cite{Govoni:2004as,2012A&ARv..20...54F}. ALP$-$photon conversion is maximised for strong magnetic fields coherent over large distances, thus galaxy cluster magnetic fields are the best places to look for signals of a CAB. Conversion is suppressed for ALP masses larger than the effective photon mass in the ICM (the plasma frequency), thus in what follows we consider only massless ALPs, but the results hold for ALP masses $m_a < 10^{-13}\text{ eV}$. This mass range is then incompatible with the ALP being the QCD Axion. The most stringent bound on $M$ for such low mass ALPs comes from astrophysics, from the lack of a gamma-ray burst coincident with the neutrino burst from SN1987a \cite{Brockway:1996yr,Grifols:1996id,Payez:2014xsa}. This bounds the inverse coupling to be
\beq
M>2\times 10^{11}\text{ GeV}.
\eeq
For a recent review of ALPs see \cite{Ringwald:2014vqa}.

In \cite{Angus:2013sua} the soft X-ray excess from the centre of the Coma cluster (radii $<700\text{ kpc}$) was studied. Using a 5 parameter, stochastic model of the magnetic field, which had been constrained by Faraday rotation observations of Coma, conversion probabilities for a range of values of the inverse coupling, $M$, and CAB energy were computed numerically. The simulated luminosity given a CAB energy density with $\Delta N_{\text{eff}}=0.5$ was then compared to the observed soft X-ray excess luminosity in the cluster. It was shown that a CAB could easily explain the observed luminosity for values
\beq
M \leq 7\times 10^{12}\text{ GeV},
\eeq
where the inequality takes into account the dependence on the energy where the CAB energy spectrum peaks (which is unknown). We show the CAB parameter space in Figure \ref{fig:coma}. The bound on $M$ from SN1987a above then puts a lower limit on the mean energy of the CAB spectrum of $\langle E_{CAB}\rangle > 40\text{ eV}$. It was also shown that the CAB hypothesis showed good agreement with the observed morphology of the soft X-ray excess, given the magnetic field model parameter uncertainties. 

In \cite{Kraljic:2014yta} this work was extended to consider the soft X-ray excess in the outer parts of the Coma cluster (up to $\sim 5\text{ Mpc}$). Using a semi-analytic method of convolving the conversion probability for a single domain with the distribution of domain sizes given by the power spectrum of the magnetic field model, it was confirmed that the soft excess in the central and outer parts of Coma could be simultaneously fit for the same CAB parameters. Not overproducing a flux in the ROSAT detector between $1-2\text{ keV}$ bounds the mean CAB energy to $\langle E_{CAB}\rangle\leq 0.37\text{ keV}$. The magnitude of the soft excess was matched for values of the inverse coupling,
\beq
5\times 10^{12}\text{ GeV}\leq M \leq 3\times 10^{13}\text{ GeV}.
\eeq
The CAB parameter space, showing the best fit regions, is plotted in Figure \ref{fig:coma}. The magnetic field at such large radii is very uncertain, the range in parameter space above allows for two extrapolations of the central magnetic field from \cite{Angus:2013sua}: one which assumes the power spectrum at large radii is the same as at low radii, and a second where the coherence lengths of the magnetic field increase with radius inversely proportional to the electron density. However since these magnetic field models are not tested against data at these radii, the bounds above are uncertain.

A CAB will convert to soft X-ray photons in the magnetic fields of all clusters, thus it is essential that the hypothesis is tested in other clusters. In this work we continue the analysis by simulating ALP$-$photon conversion in the clusters A665, A2199 and A2255. We do this by simulating cluster magnetic fields and calculating the expected soft X-ray flux from CAB conversion to photons in the three clusters. For the cluster magnetic fields, we assume they are well represented by the 5 parameter stochastic, magnetic field model of \cite{Murgia:2004}. These models have been produced for the three clusters we look at, and have had their model parameters constrained by using Faraday Rotation measurements and synchrotron radio emission simulation \cite{2006AA...460..425G,Vacca:2010ss,2012AA...540A..38V}. In addition, all three clusters were analysed as part of the 38 cluster survey \cite{astroph0205473}.

The cluster A665 shows no evidence for a soft X-ray excess, whereas A2199 has hints for a soft X-ray excess at very low significance, and A2255 has a soft X-ray excess, though again at low significance. The test of the CAB against the lack of observation in the A665 is an important one because a large proportion of clusters have no observed excess. Since a CAB will always produce a soft excess, it is important to check that the excess is not predicted to be observably large for clusters where no excess is seen. 

Another important check is to confirm whether the best fit CAB parameters agree between clusters. We find the best fit parameter space in A2199 and A2255 by scanning over the CAB parameters and matching to the observed soft X-ray excesses. We can use this to then show that the CAB parameter space agrees between the four clusters we have studied so far. We also compare the observed and simulated soft excess morphology in the clusters A2199 and A2255.

The outline of the paper is as follows. In the next section, Section \ref{sec:excess}, we review soft X-ray excess observations with particular focus on the three galaxy clusters of interest to this study. In Section \ref{sec:method} we detail the numerical methodology used to simulate the magnetic field, calculate conversion probabilities and compare to data. In Section \ref{sec:results} we state the results for the three clusters and compare the parameter space for the four clusters studied so far. We conclude in Section \ref{sec:conclusions}.

\section{The Cluster Soft X-ray Excess}
\label{sec:excess}
In this section we give a brief review of cluster soft X-ray excess observations, focusing on the three clusters A665, A2199 and A2255. For a comprehensive review of the subject we refer the reader to \cite{08010977}---see also the review in \cite{Angus:2013sua} which includes a full discussion of the potential astrophysical explanations.

The cluster soft X-ray excess is a phenomenon which has been observed in a large number of galaxy clusters. The excess is seen above the thermal emission from the hot, ionised intra-cluster medium. Clusters are keV temperature objects, thus thermally emit across the X-ray regime. The soft X-ray emission from a cluster can be predicted using observations at higher X-ray energies, in many clusters the observed soft X-ray emission is in excess of this prediction. The observed excess is diffuse, extends to large radii, and is in the extreme ultraviolet (EUV, $\sim 100\text{ eV}$) and soft X-ray ($\sim 0.25\text{ keV}$) bands. It has been observed in a large number of clusters with the EUVE and ROSAT satellites, and in a small number with the BeppoSAX, XMM-Newton, Suzaku, and Chandra satellites. Its existence has been a subject of controversy for many years, with several analyses disagreeing about its existence in specific clusters. Its origin has also been a subject of controversy. Proposed astrophysical explanations, such as thermal emission from a second, cooler component of the ICM or in the form of a warm-hot intergalactic medium (WHIM), or non-thermal emission from inverse Compton scattering of CMB photons, run into difficulty and are constrained by other observations.

The cluster soft X-ray excess was first observed in the Virgo and Coma clusters in 1996 with the EUVE satellite \cite{Lieu1996a,Lieu1996b,Bowyer96}. It was subsequently seen in many other clusters with EUVE, although analyses questioning the background subtraction in many of these clusters leaves the status of observations of the cluster soft X-ray excess (except for Virgo and Coma where the excess is established) with EUVE unclear. The excess has also been observed in a large proportion of clusters definitively with the ROSAT X-ray satellite. In a sample of 38 galaxy clusters analysed with ROSAT data \cite{astroph0205473}, a third of the clusters were found to host significant, diffuse, excess emission in the $0.1-0.28\text{ keV}$ band, revealing the excess to be a widespread phenomenon. Also, in this study, upon dividing clusters into concentric annular regions, it was observed that the soft excess was preferentially seen at large radii.

Observations with newer satellites such as XMM-Newton, Suzaku and Chandra have also revealed excesses in some clusters. These satellites have much better spectral resolution than EUVE and ROSAT, however due to the small fields of view of these satellites the subtraction of the temporally and spatially varying soft X-ray background is challenging. The newer satellites also suffer from soft proton flares which produce a background of events indistinguishable from soft X-ray photons. Thus the existence of the soft excess in observations with these satellites is unclear. Also because of this there is little spectral information available about the excess.

\begin{figure}[t]
\centering
	\subfloat[A665]{
		\includegraphics[width=0.48\textwidth]{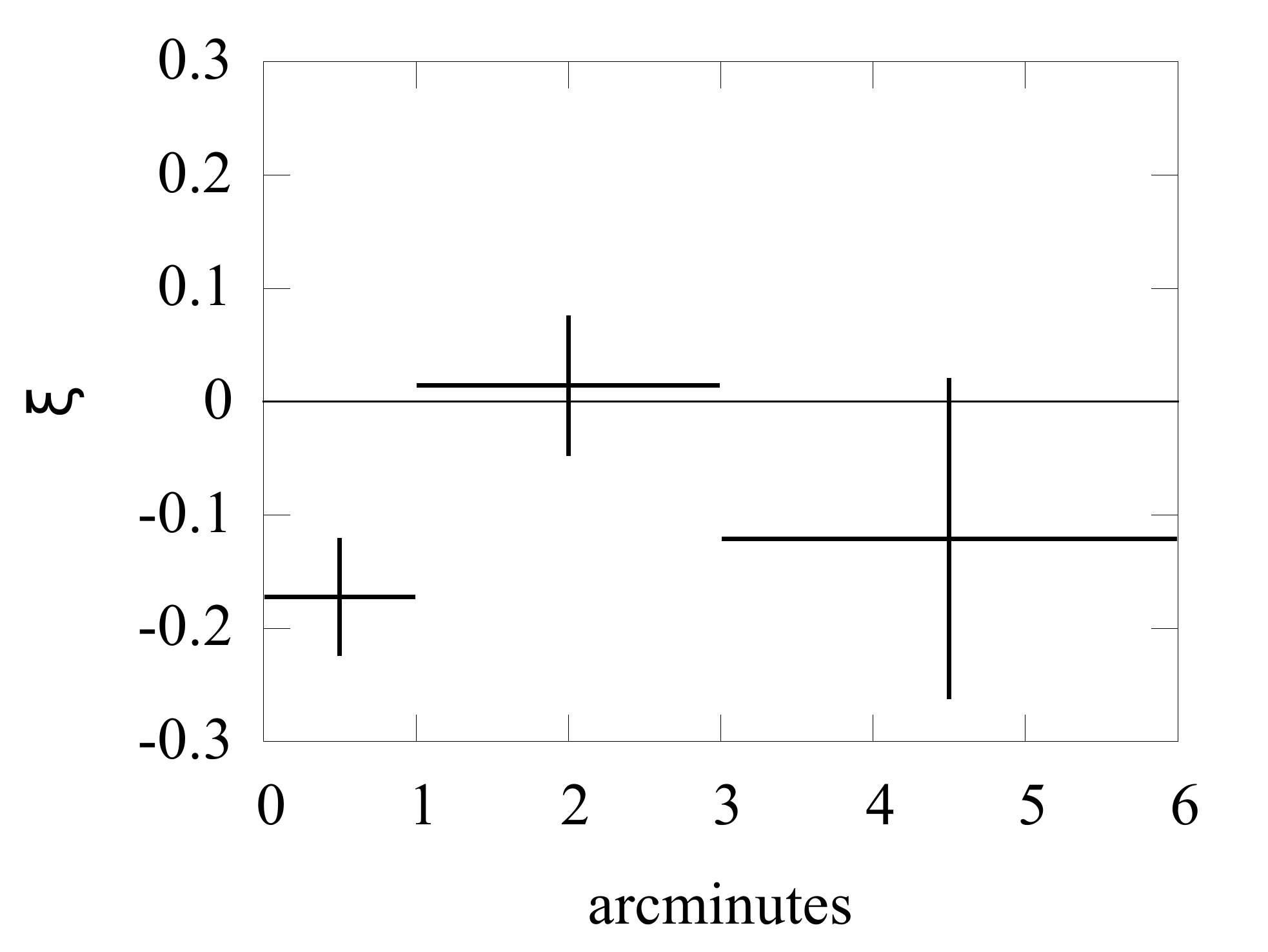}
	}
	\subfloat[A2199]{
		\includegraphics[width=0.48\textwidth]{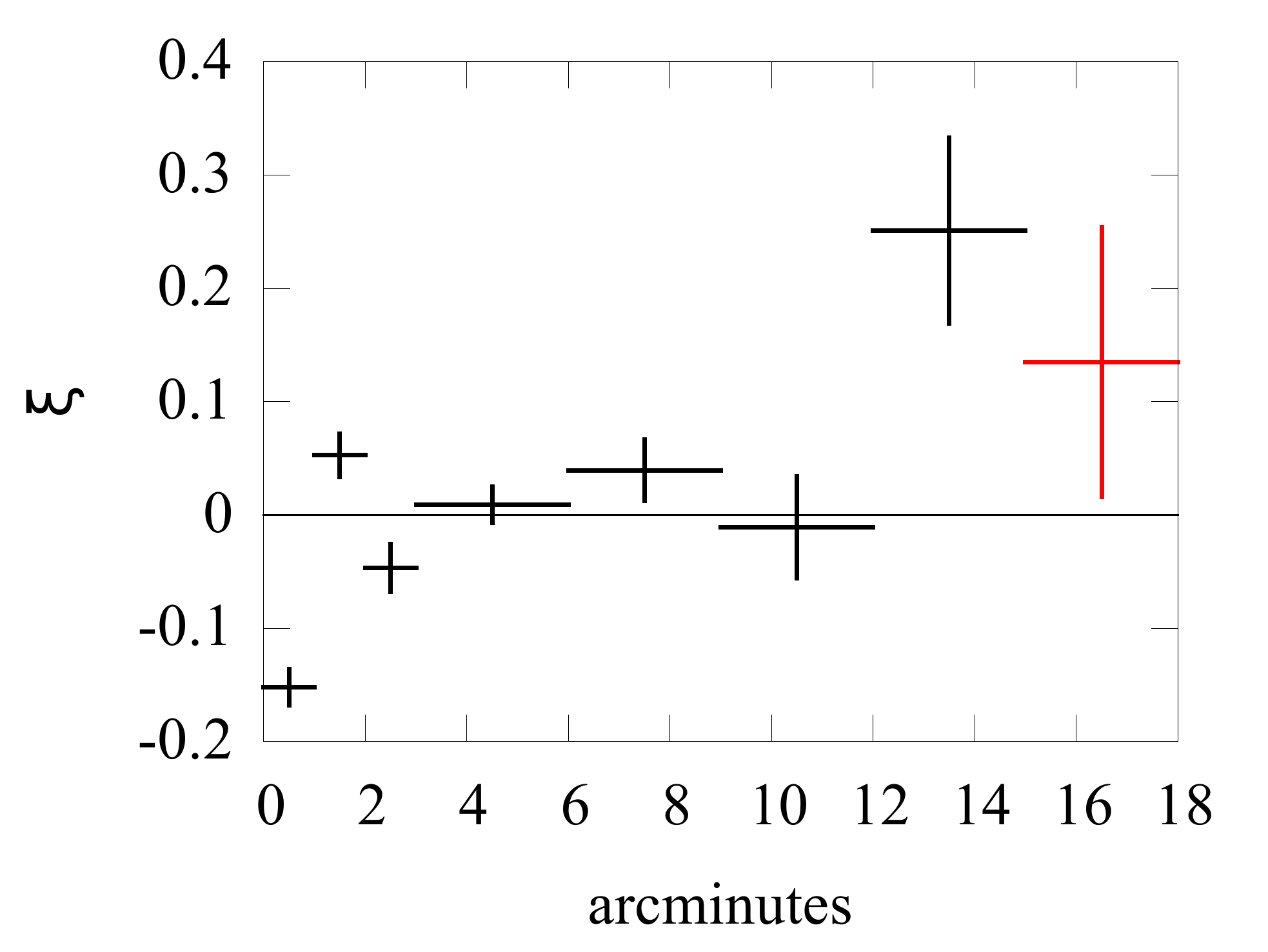}
	}
	\\
	\subfloat[A2255]{
		\includegraphics[width=0.48\textwidth]{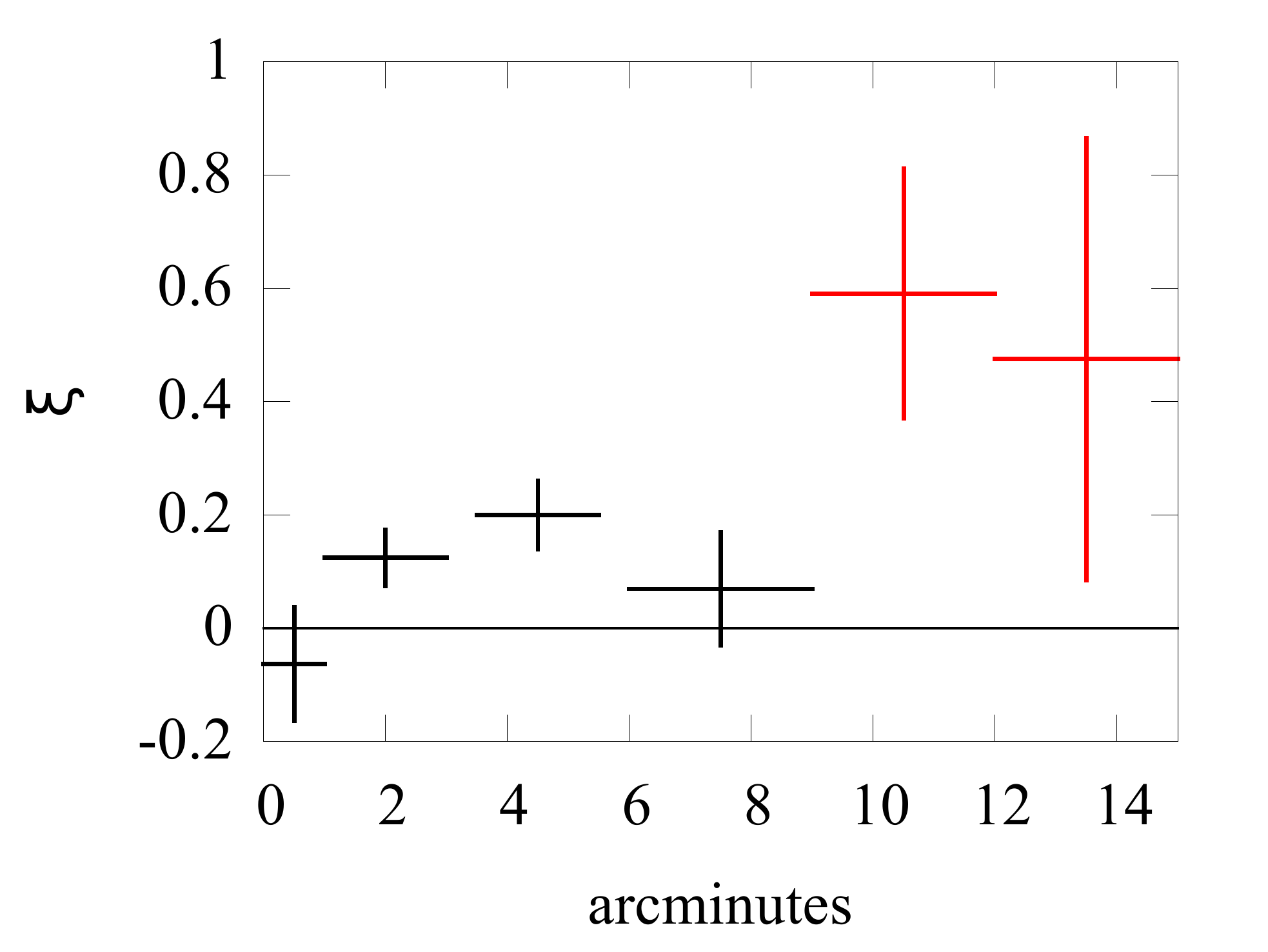}
	}
	\caption{The fractional excess, $\xi$, over the expected thermal bremsstrahlung emission in the $0.1-0.28\text{ keV}$ band, in several annuli for the three clusters a) A665, b) A2199, and c) A2255. Data taken from \cite{astroph0205473}. Points coloured red indicate that the total (thermal and soft excess) flux in the $0.1-0.28\text{ keV}$ band does not exceed the background by 25\%.}
	\label{fig:obs}
\end{figure}

Throughout this work we will compare our simulations to the observations of \cite{astroph0205473}, which is the largest complete study of the soft X-ray excess in 38 galaxy clusters, and currently appears unchallenged. The excess is usually phrased as the fractional excess, $\xi$, which is defined as
\beq
\xi\equiv\frac{p-q}{q},
\eeq
where $p$ and $q$ are the observed and expected counts in the $0.1-0.28\text{ keV}$ band respectively. We show the fractional excess in several concentric annuli for the three clusters involved in this study in Figure \ref{fig:obs}.

\subsubsection*{A665}

ROSAT observations of the galaxy cluster Abell 665 in \cite{astroph0205473} find no excess. Indeed the inner part of the cluster seems to show a deficit of soft X-rays which is taken as a sign of some cooler gas resulting in absorption of the soft X-ray component of the ICM. Observations at larger radii show results consistent with zero excess counts. No other analyses of observations of A665 have been performed, that we are aware of.

\subsubsection*{A2199}

The second galaxy cluster we consider is Abell 2199. Many analyses of the soft X-ray excess in A2199 have been performed. An excess in A2199 has been observed with three satellites: the EUVE satellite and the X-ray satellites BeppoSAX and ROSAT \cite{astroph9902300,astroph9911001,astroph9910298,1999ApJ...519L.119K,2000A&A...364..497L,astroph0204345,2002ApJ...565L..17B,2002ApJ...574L...1K}. However due to controversies about the background subtraction in EUVE and BeppoSAX, the status of observations of the soft excess in A2199 with these satellites is unclear.

Averaging over the whole of the cluster, the ROSAT analysis of \cite{astroph0205473} found no compelling evidence for a soft X-ray excess. Excluding the large deficit at $<1\text{ arcminute}$, there is small overall excess, however due to the large scatter in the data it has very low statistical significance.

\subsubsection*{A2255}

Observations of Abell 2255 in \cite{astroph0205473} revealed an excess, though averaged over the cluster it is still at low significance.
Although the excess is not as statistically significant as the Coma cluster, an excess of $\sim 20$\% was seen between 1 and 6 arcminutes from the cluster centre. An even larger excess of $\sim 50$\% was seen between 9 and 15 arcminutes, though the total signal (thermal emission from hot ICM plus any excess emission) over background in this region was too poor to draw conclusive results and were thus not analysed further. This is the only analyses of the soft X-ray excess that has been performed on A2255 that we know of.

\section{Numerical Methodology}
\label{sec:method}
In the presence of an external magnetic field, photons and ALPs can oscillate into one another in a way analogous to neutrino oscillations. In galaxy clusters this external magnetic field is supplied by the magnetised, turbulent ICM. Here we give an overview of the power-law, stochastic field model we assume for the three clusters' magnetic fields. The magnetic field model is a multi-parameter model, and we detail how the parameters have been constrained for each of the clusters in the next sub-section. Then we solve the coupled equation of motion for the photon-ALP system both analytically for a single homogeneous domain, and numerically for the discrete magnetic field model we use. We include brief details of the numerical methodology, for more details we refer the reader to \cite{Angus:2013sua}, where the computational details are made more explicit. We end the section by detailing the method used to compare our results to data.

\subsection{Magnetic Field Model}
\label{sec:Bmodel}
Galaxy clusters are known to be magnetised, see \cite{Govoni:2004as,2012A&ARv..20...54F} for reviews of observations and methods to measure cluster magnetic fields. Observations in radio waves have shown that a large number of clusters contain large areas of diffuse radio emission (radio halos/relics) which cannot be attributed to known sources. Such radio structures are produced by synchrotron emission of a relativistic electron population in the magnetic field of the cluster. Observations of polarised sources in and behind galaxy clusters also exhibit Faraday rotation consistent with treating the cluster's magnetic field as a `Faraday screen', whose magnetic field rotates the plane of polarisation of the radio emission by making the intra-cluster medium birefringent.

Detailed measurement of the magnetic field in clusters is challenging. Synchrotron radio emission is degenerate between parameters of the relativistic electron population and the magnetic field, and the properties of neither of these are known independently. The Faraday rotation observations are limited by the number of polarised radio sources behind an individual cluster and only depend on the integral of the parallel component of magnetic field along the line of sight. Thus model assumptions need to be made to constrain the magnetic field. Equipartition arguments---a minimum energy criterion that the cosmic ray and magnetic field energy densities should be the same---give order of magnitude estimates for the magnetic field strength. Since Faraday rotation measurements are related to the line of sight integral of the magnetic field, if one makes an assumption that the magnitude of the magnetic field is constant along many single-sized domains, the variance of the Faraday rotation measures gives the size of these domains, which is a hint to the typical coherence length of the field along the line of sight. Using maps of the induced Faraday rotation can similarly give hints to the typical size of the coherence lengths of the magnetic field. More in depth analyses involve simulating a magnetic field, given a specific model assumed for the field, and using the observed radio halo and Faraday rotation data to constrain the parameters of the model. Using the power spectrum of the Faraday rotation is another method used to constrain the magnetic field power spectrum. However, it has been argued that the use of Faraday rotation measurements does not take into account the intrinsic rotation in the source of the polarised signal and thus is ill-suited to constraining the cluster-wide magnetic field \cite{2003ApJ...588..143R}.

In this study we assume the cluster magnetic fields are well represented by the model first proposed in \cite{Murgia:2004}. The simulated magnetic field is a stochastically-generated, Gaussianly-distributed field that is tangled over a range of scales, and has a power-law power spectrum. The magnitude of the magnetic field is assumed to decrease with radius as some power of the thermal electron density of the ICM. Such fields have shown good fits to Faraday rotation measurements and synchrotron radio halos in many clusters \cite{Murgia:2004,2006AA...460..425G,10020594,Vacca:2010ss,2012AA...540A..38V,13057228}. However, we caution that it is an idealised model and thus may not capture the full details of the magnetic fields in these clusters. This magnetic field model was used to study the soft X-ray excess in both the centre and the outskirts of the Coma galaxy cluster \cite{Angus:2013sua,Kraljic:2014yta}. Here we summarise its key features.

We first generate the Fourier-space vector potential randomly from a Rayleigh distribution with power law scaling
\begin{equation}
|A_k|^2 \sim k^{-n},
\end{equation} 
and uniformly-distributed phase. The Fourier-space magnetic field, given by $\tilde{B}(k)=ik\times \tilde{A}(k)$, is then Fourier transformed back to position space to generate a Gaussianly-distributed, isotropic, tangled, divergence-free magnetic field, with power spectrum
\begin{equation}
4\pi k^2|B_k|^2\sim k^{-\zeta},
\end{equation}
where $\zeta=n-4$. The Fourier vector potential is set to zero outside of the momentum range $k_{min}$ to $k_{max}$, corresponding to physical scales $\Lambda_{max}=2\pi/k_{min}$ and $\Lambda_{min}=2\pi/k_{max}$. This magnetic field is then modulated by multiplying by some function $f(n_e)\propto {\left(n_e\right)}^{\eta}$, and normalised such that the core of the cluster has a magnetic field of magnitude $B_0$, such that
\beq
B(r)=B_0\left(\frac{n_e(r)}{n_0}\right)^{\eta}.
\eeq
We take time here to point out there are two interesting values for $\eta$: the first is the case where the magnetic field energy density ($\varepsilon_B\propto B^2$) falls with the electron energy density, i.e. $B^2\propto n_e$ and $\eta=0.5$; the second is the case where the magnetic field lines are `frozen-in' to the plasma,\footnote{Briefly, the magneto-hydrodynamical ICM is in a regime where the magnetic Reynolds number is large (corresponding to turbulence). In this regime the magnetic field lines `move with the plasma', specifically this regime corresponds to $B\cdot A=\text{const}$ (with A the cross-sectional area) and hence $B\propto n_e^{2/3}$.} which corresponds to the case $\eta=2/3$. 

The electron/gas density of clusters follows an isothermal $\beta$-model profile
\beq
n_e(r)=n_0\left(1+\frac{r^2}{r_c^2}\right)^{-\frac{3}{2}\beta},
\eeq
with $n_0$ the central electron density and $r_c$ the core radius of the cluster. In many clusters the core has cooled (such as A2199), leading to a spike in the electron density in the centre \cite{09110409}. The electron density profile in these cool-core clusters then usually follows a double $\beta$-model, which is the sum of two $\beta$-models,
\beq
n_e(r)=n_0\left(1+\frac{r^2}{r_c^2}\right)^{-\frac{3}{2}\beta}+n_{0,\text{cool}}\left(1+\frac{r^2}{r_{c,\text{cool}}^2}\right)^{-\frac{3}{2}\beta_{\text{cool}}},
\label{eq:doublebeta}
\eeq
with $r_{c,\text{cool}}\ll r_c$ and typically $n_{0,\text{cool}}\geq n_0$.

The magnetic field model then has 5 parameters: the power spectrum index $\zeta$, the maximum and minimum lengthscales the magnetic field is `tangled' over $\Lambda_{max}$ and $\Lambda_{min},$\footnote{Note that the scale $\Lambda$ corresponds to the wavelength of a full period of magnetic field `oscillation', the field will thus be coherent over a scale $\lesssim\Lambda/2$.} the radial scaling of the magnetic field $\eta$, and the magnetic field magnitude in the centre of the cluster $B_0$. These parameters are constrained using Faraday rotation maps or synchrotron radio halos, and are listed in Table \ref{tab:parameters}, along with the $\beta$-model parameters.

The magnetic field is simulated on a $1000^2\times2000$ grid, where 2000 is the number of points along the propagation direction. The conversion probabilities are computed on a $1000\times 1000$ grid where, to get maximal radial coverage, the centre of the $\begin{pmatrix}x,&y\end{pmatrix}$ plane is taken at a corner of the grid. By the assumed symmetry of the cluster we are simulating---i.e. the spherically-symmetric electron density and magnetic field profiles---the same conversion probabilities are assumed to hold in the three other quadrants of the $\begin{pmatrix}x,&y\end{pmatrix}$ plane. The ALPs are propagated numerically by assuming that the magnetic field and electron density are constant between grid points.

\subsubsection*{A665}

Here we summarise the method and the results of \cite{Vacca:2010ss} in constraining the  above model parameters in the A665 galaxy cluster. The parameters for A665 are constrained by simulating mock radio halo images, upon assuming the form of the relativistic electron population, and comparing to the observed radio halo. The analysis proceeds in two steps: first, equipartition arguments are used to determine the radial scaling of the magnetic field, the parameter $\eta$. Secondly mock radio halos are produced and compared to data to constrain the field strength and power spectrum.

In the first step the electron population is taken as a power law where the spectral index of the electron population is related to the spectral index of the observed radio halo. The number of relativistic electrons in the cluster is set point-by-point such that at that point there is equipartition between the relativistic electrons and magnetic field. We note that the spatial form for the relativistic electron population is an assumption, since little is known about it or how it is produced. Equipartition is the assumption that the energy stored in the magnetic field and in the cosmic ray electrons is equal. Assuming the magnetic field strength is proportional to the electron density to some power $\eta$ the radial profile of this spherically symmetric radio halo is predicted as a function of $\eta$. This is compared to the radial profile of the observed radio halo, giving a best fit value of $\eta = 0.47\pm 0.03$, which we note is very close to the interesting value mentioned earlier, where the energy density of the magnetic field falls proportional to the thermal electron energy density.

A full 3D magnetic field is produced using the method described earlier. The same form for the electron population as used in the first step is used to simulate mock radio halos for this field. The predicted radio emission from the idealised, spherically-symmetric radio model used in the first step is subtracted from both the observed radio halo and the simulated radio halo. The RMS of the residuals for the simulated halos are then plotted as a function of $\Lambda_{max}$, and show a clear trend that they increase with $\Lambda_{max}$. The simulated RMS matches the observed radio halo's for $\Lambda_{max}\sim 450$ kpc.

On top of this there is a more qualitative result that there is a clear change in shape of the radio halo when going from small to large $\Lambda_{max}$, at larger $\Lambda_{max}$ the halo becomes more anisotropic and also there is a clear separation between the centre of the ICM plasma and the centre inferred from radio observations. One can plot this offset distance for the different values of $\Lambda_{max}$, there is a clear trend for larger offsets for larger $\Lambda_{max}$ values. The value where the simulations match the offset of the observations is slightly higher, at roughly 500 kpc.

\subsubsection*{A2199}

The parameters of the model for the magnetic field of A2199 have been constrained by producing mock Faraday rotation images and comparing to the observed Faraday rotation in a radio source located at the centre of the cluster \cite{2012AA...540A..38V}. All Faraday rotation data is contained within $20\text{ kpc}$ of the centre of the cluster, thus in our analysis we are extending the magnetic field beyond the range which it is known to be valid. The X-ray brightness of the cluster is not completely spherically-symmetric, it contains two elliptical regions to the east and the west of the cluster where the electron density drops significantly. The simulations of \cite{2012AA...540A..38V} showed that there was very little effect on results when they were included as when they were left out. Since we are averaging the probabilities over concentric annuli, and since the features are located within $20\text{ kpc}$ and we simulate out to $>500\text{ kpc}$, we do not expect these regions to change the analysis in any appreciable way, we thus do not include these elliptical areas in our simulation. 

A2199 is a cool-core cluster and thus the gas density follows a double $\beta$-model, see Equation \ref{eq:doublebeta}. The electron density in the cool-core is $\sim 30$ times that in Coma and A665 for instance. The cool-core electron density spike drops rapidly beyond 9 kpc, whereas the normal ICM electron population has a core radius of 26 kpc. Again since we simulate out to $\sim 400\text{ kpc}$ we do not expect the cool-core to have a big effect on the results.

The magnetic field was simulated scanning over all five parameters of the model, using the power spectrum of the Faraday rotation to constrain the magnetic field power spectrum parameters, and maps of Faraday rotation to constrain the radial scaling. The parameters of the magnetic field of A2199 are not well constrained from this analysis.  The parameters $B_0$ and $\eta$ are degenerate with respect to Faraday rotation measurements, i.e. the same average magnetic field can be obtained by reducing (increasing) $B_0$ whilst simultaneously reducing (increasing) $\eta$ such that the field falls off less (more) rapidly with radius. The resulting values of the central magnetic field and radial parameters are $B_0=11.9\pm 9$ and $\eta=0.9\pm 0.5$. There is a similar degeneracy between the maximum lengthscale the field is tangled over and the power spectrum of the field, large spectral indices put more power at large scales, increasing $\Lambda_{max}$ has the same effect. Thus again these parameters are poorly constrained, although we keep these parameters fixed at their central values in these simulations.

\subsubsection*{A2255}

The magnetic field parameters in A2255 are constrained in \cite{2006AA...460..425G}, using polarised emission from three galaxies. In this cluster a synchrotron radio halo is simulated, as well as mock Faraday rotation maps. The radial scaling of the magnetic field is constrained in two ways, firstly the root mean square of the Faraday rotation measures in each galaxy should be proportional to the X-ray brightness of the ICM, to some power. This poorly constrains $\eta$ to anywhere between $-0.5$ and $1$. Secondly there is the observation that the X-ray brightness profile follows the radio halo brightness profile thus it is assumed that the populations of thermal and non-thermal electrons, and hence also the magnetic field (by equipartition) follow the same profile. Thus it is assumed in their simulations that $n_e \propto B^2$. In these simulations we will consider two values: $\eta=0.5$ and $\eta=0.7$, which (roughly) correspond to the two interesting cases mentioned earlier.

It was shown that the Faraday rotation is best fit for a magnetic field that has increasing power spectrum index as a function of radius. The best fit global magnetic field is one in which the magnetic field has $\zeta=0$ for $r<r_c$ and then an exponential fall off beyond $r_c$, and then a non-Gaussian $\zeta=2$ magnetic field with larger $\Lambda_{min}$ for radii beyond the core radius. In our simulations we use this model, but assume both parts of the magnetic field follow a Gaussian distribution.

Larger spectral indices result in more power residing in larger lengthscales, and thus the typical coherence lengths of the field are larger. Thus this choice of a rising spectral index with radius corresponds to magnetic field coherence lengths increasing with radius. This sort of behaviour is expected on general theoretical grounds: as the electron density decreases, the typical lengthscale of the problem $\propto 1/n_e^{1/3}$ gets larger. Such behaviour of the magnetic field was considered for ALP$-$photon conversion in the outskirts of Coma in \cite{Kraljic:2014yta}. Though the model of the field as a `stitching' together of two fields with differing spectral index with a smoothing function between them is unrealistic, it is nevertheless a reasonable test model to study the CAB in A2255.

\begin{table}[t]
	\centering
		\begin{tabular}[t]{c|c|c|c|c|c}
			\hline
			 & $\quad\ $A665 \cite{Vacca:2010ss}$\ \quad$ & \multicolumn{2}{c|}{A2199 \cite{2012AA...540A..38V}} & \multicolumn{2}{c}{$\quad$ A2255 \cite{2006AA...460..425G} $\quad$ } \T \B\\
\hline
\hline
			& & $\ $Cool-core$\ $ & $\ \, $Hot ICM$\,\ $  & $\ \ $Inner$\ \ $ &$\,$ Outer$\,$ \T \B \\
			\hline
			$r_c$ (kpc) & 340 & 9 & 25 & \multicolumn{2}{c}{432}\T\B\\
			$\beta$ & 0.76  & 1.5&0.39& \multicolumn{2}{c}{0.74}\T\B\\
			$\quad n_0$ ($10^{-3}$ cm$^{-3})\quad $& 3.44 & 74&27& \multicolumn{2}{c}{2.20} \T\B\\
			\hline
			$\zeta$ & 5/3& \multicolumn{2}{c|}{0.6} & 0 & 2 \T\B\\
			$\Lambda_{min}$ (kpc)& 4 & \multicolumn{2}{c|}{0.7}& 4& 64\T\B\\
			$\Lambda_{max}$ (kpc)& 450 & \multicolumn{2}{c|}{35} & 1000 & 1000\T\B\\
			$B_0$ ($\mu$G)& 1.3&  \multicolumn{2}{c|}{$6.0-11.9$} & \multicolumn{2}{c}{2.5} \T\B\\
			$\eta$& 0.5 & \multicolumn{2}{c|}{$0.5-0.9$} & \multicolumn{2}{c}{$0.5-0.7$}\T\B\\
			\hline
	\end{tabular}
	\caption{Magnetic field and $\beta$-model parameters for the three clusters considered. The A2255 Inner and Outer columns refer to the magnetic field within and without the core radius respectively. The A2199 cool-core and hot ICM columns refer to the parameters for the two $\beta$-models which make up the cool-core and the normal hot ICM respectively.}
\label{tab:parameters}
\end{table}

\subsection{Conversion Probabilities}

ALPs can convert into photons via the Lagrangian in Equation \ref{eq:alpcoupling}, the conversion is governed by the inverse coupling $M=g_{a\gamma\gamma}^{-1}$,
\beq
\mathcal{L}\supset \frac{1}{4}g_{a\gamma\gamma}\,a\,F\tilde{F}=\frac{1}{M}a\,\vec{E}\cdot\vec{B}.
\eeq
Due to this coupling, in an external magnetic field, one can write the system of the ALP and photon as a coupled three-body problem of the ALP and the two polarisation states of the photon (parallel and perpendicular to the magnetic field). The ALP$-$photon system then oscillates in an analogous way to neutrinos. The system has a non-diagonal linearised equation of motion \cite{Raffelt},
\begin{equation}
\label{eq:EqofMotion}
\left(\omega + \left(\begin{array}{c c c}
			\Delta_{\gamma} & 0 & \Delta_{\gamma a x} \\
			0 & \Delta_{\gamma} & \Delta_{\gamma a y} \\
			\Delta_{\gamma a x} & \Delta_{\gamma a y} & \Delta_{a}
		   \end{array}\right) - i\partial_z\right)\left(\begin{array}{c}
								\perpphoton \\
								\parphoton \\
								\axion
							      \end{array}\right)= 0,
\end{equation}
where $\omega$ is the ALP/photon energy. The photon dispersion relation is altered by the ICM and thus the photon picks up an effective mass of the plasma frequency of the ICM $\omega_{pl}^{\,2}=4\pi\alpha n_e/m_e$, giving the diagonal mass term for the photon states $\Delta_{\gamma} = -\omega_{pl}^2/2\omega$. The ALP has the diagonal mass term $\Delta_a = -m_a^2/\omega$, we set this term to zero in the subsequent analysis. The two off-diagonal terms which induce the mixing between the ALP and the photon, are $\Delta_{\gamma a i} = B_i/2M$. We have set to zero the mixing between the two photon polarisation states caused by Faraday rotation as this will not affect ALP conversion to photons.

The wavefunction leaving the cluster is thus given by
\begin{equation}
\label{eq:homsoln}
\left(\begin{array}{c}
	\perpphoton \\
	\parphoton \\
	\axion \\
\end{array}\right) =  \exp \left(-i \int \left(\begin{array}{c c c}
							\Delta_{\gamma}(z) & 0 & \Delta_{\gamma a x}(z) \\
							0 & \Delta_{\gamma}(z) & \Delta_{\gamma a y}(z) \\
							\Delta_{\gamma a x}(z) & \Delta_{\gamma a y}(z) & \Delta_{a}(z)
		   				\end{array}\right) \dd z\right) \left(\begin{array}{c}
				\perpphoton \\
				\parphoton \\
				\axion
		      \end{array}\right)_0,
\end{equation}
where the subscript 0 denotes the original wavefunction, which for our purposes is a pure ALP state.
Since the magnetic field is generated on a discrete grid, we split the integral into a discrete sum allowing us to iteratively `propagate' the wavefunction from one grid point to the next:
\begin{equation}
\left(\begin{array}{c}
	\perpphoton \\
	\parphoton \\
	\axion \\
\end{array}\right)_{n+1} =  \exp \left(-i \left(\begin{array}{c c c}
							\Delta_{\gamma,n} & 0 & \Delta_{\gamma a x,n} \\
							0 & \Delta_{\gamma,n} & \Delta_{\gamma a y,n} \\
							\Delta_{\gamma a x,n} & \Delta_{\gamma a y,n} & \Delta_{a,n}
		   				\end{array}\right) \Delta z\right) \left(\begin{array}{c}
				\perpphoton \\
				\parphoton \\
				\axion
		      \end{array}\right)_n,
\end{equation}
with $\Delta z$ the unit cell size.

Practically, we rotate the fields at each grid point such that the matrix is diagonal, the wavefunction is then `propagated' to the next grid point where we rotate back to the original basis. We repeat this step iteratively until reaching the edge of the cluster. The conversion probability is then the sum of the square of the photon components of the wavefunction.

The result is a grid of conversion probabilities, which we compute for several ALP energies between $25\text{ eV}-1\text{ keV}$. This grid is divided into ten concentric annuli and the conversion probability in each annulus is averaged. In this way we end up with an ALP$-$photon conversion probability distribution which is a function of radius and energy, $ P (r,E)$, which can be used to calculate the luminosity over a given energy and radial range.

\subsubsection*{Single Domain Solution}

Let us pause for a moment and consider the analytic solution for an ALP propagating through one magnetic field domain with constant field and electron density. The probability that an ALP converts into a photon is then simply given by
\begin{equation}
\label{eq:convprob}
	P(a\rightarrow\gamma)=\sin^2(2\theta)\sin^2\left(\frac{\Delta}{\cos 2 \theta}\right),
\eeq
where $\tan 2 \theta = \frac{2B_{\perp}\omega}{M m_{eff}^2}$, $\Delta=\frac{m^2_{eff}L}{4\omega}$, $m_{eff}^2=m_a^2 -\omega_{pl}^2$, $\omega$ is the ALP energy, $L$ is the size of the domain, $\omega_{pl}$ is the plasma frequency defined above, and $B_{\perp}$ is the component of the magnetic field perpendicular to the propagation direction. The mixing is thus governed by two dimensionless parameters (called angles): $\theta$ and $\Delta$, given by
\beq
	\theta \approx \frac{B_{\perp}\omega}{M m_{eff}^2} = 5.6\cdot 10^{-4}\left(\frac{10^{-3}\hbox{ cm}^{-3}}{n_e}\right)\left(\frac{B_{\perp}}{2\mug}\right)\left(\frac{\omega}{200\hbox{ eV}}\right)\left(\frac{10^{13}\hbox{ GeV}}{M}\right),
\label{eq:theta}
\eeq
\beq
	\Delta=2.7\left(\frac{n_e}{10^{-3}\hbox{ cm}^{-3}}\right)\left(\frac{200\hbox{ eV}}{\omega}\right)\left(\frac{L}{10\kpc}\right).
\label{eq:delta}
\eeq
Where as discussed earlier, we have taken $m_a=0$. To get an estimate of the conversion probabilities, we take an approximation known as the small-angle regime $\theta \ll 1$, $\Delta \ll 1$, which is always true for $\theta$, but not true for $\Delta$ for large values of $n_e$ (and hence small radii), large $L$, or low ALP energies. Then the probability an ALP will convert to a photon over $D/L$ domains, with $D$ the full cluster size is simply
\beq
	P(a\rightarrow\gamma) = 0.9\cdot 10^{-3}\left(\frac{D}{1\text{ Mpc}}\frac{L}{10\text{ kpc}}\right) \left(\frac{B_{\perp}}{2\mug}\frac{10^{13}\hbox{ GeV}}{M}\right)^2.
\label{eq:smallangle}
\end{equation}
We thus see that clusters can be extremely efficient at ALP$-$photon conversion.

While the above discussion is illuminating, it can only give us an order of magnitude estimation for the typical conversion probability in clusters. A constantly varying, turbulent, multi-scale magnetic field requires a full numerical calculation to calculate the conversion probabilities. However, it was shown in \cite{Angus:2013sua} that the conversion probabilities from the full simulation can be understood well by looking at the single domain conversion formula. Typically there are two regimes, one where the angle $\Delta$ is large (low ALP energies, or high electron densities), and one where it is small and the full small angle regime formula can be used. In the large-$\Delta$ regime
\beq
P(a\rightarrow\gamma)\propto\frac{\omega^2}{M^2}\left(\frac{B(r)}{n_e(r)}\right)^2\left(\frac{D}{L}\right).
\label{eq:largedelta}
\eeq
Thus in this regime the conversion probabilities are energy dependent, and behave with radius as $P\propto n_e^{\ 2(\eta-1)}$. This will (for $0<\eta<1$) produce conversion probabilities which increase with radius. For small-$\Delta$, we see in Equation \ref{eq:smallangle} that the probability will fall with radius, $P\propto n_e^{\ 2\eta}$, and that the conversion is no longer dependent on the energy.

\subsection{Luminosity and Fractional Excess}
\label{sec:lum}

We use the ALP$-$photon conversion probability distribution discussed earlier, which is a function of radius and ALP energy, to compute the soft X-ray luminosity from CAB conversion in the cluster. We do this by integrating the product of the probability function $P(r,E)$ and the CAB energy spectrum $E\, \frac{\dd N}{\dd E}$ over the area of the annulus of interest and over the energy interval $0.1-0.28\text{ keV}$. The CAB energy density is normalised relative to the CMB by setting $\Delta N_{\text{eff}} = 0.5$.

The analysis of \cite{astroph0205473} lists the expected and observed count rates in the R1 and R2 bands (R1R2, $0.1-0.28\text{ keV}$) of the ROSAT PSPC detector. In the cases where an excess is seen, the luminosity in the $0.2-0.4\text{ keV}$ band is computed, but note that the luminosity is missing in a large number of cases, specifically in the case of A665 where no excess is detected. Thus to compare to these observations we use the NASA PIMMS software \cite{PIMMS}, we use this to predict the count rate in the ROSAT PSPC R1R2 band, given a flux and spectral shape.

As can be seen in \cite{astroph0205473} the assumed spectral shape of the emission can have a large impact on the luminosity for a given a count rate. This is both due to the different energy channels of the input and output, and because the detector response varies over the R1R2 band. To use the PIMMS tool we need to state what spectral shape the signal has---which would be a modified CAB spectrum to take into account the energy dependence of the conversion probabilities. Since we are converting between luminosity and count rate in the same energy channel, we checked to see the importance of the specific spectral model used. We use two models: a thermal blackbody spectrum which peaks at $150\text{ eV}$, and a single temperature ($0.08\text{ keV}$) thermal bremsstrahlung MEKAL plasma model, which was one of the models used in \cite{astroph0205473}, and was the model used in \cite{Angus:2013sua}. We found little difference between the count rate prediction for the two models. The count rate is also slightly sensitive to the temperatures of both the blackbody and the plasma model (which in principle should be changed when scanning over the mean CAB energy parameter later), the largest deviations only came when the temperatures were pushed much higher or lower than the values above, and thus we keep these temperatures fixed.

The largest uncertainty comes from the count rate prediction itself. There is likely an inherent systematic uncertainty in simulating the count rate in the ROSAT detector for a given luminosity. For instance, we took the excess luminosities given in \cite{astroph0205473} and converted them to count rates in the ROSAT R1R2 band using PIMMS to compare with observed count rates. We found that the count rates predicted by PIMMS were factors $0.5-2$ different than the observed counts, though these errors are probably larger due to the different energy ranges of the luminosity and count rates mentioned earlier. Thus it is important that we do not claim to be too precise with the count rate prediction from the simulation, as such we allow the count rates to vary by $\pm 50\%$ when constraining the CAB parameter space later. 

\section{Results}
\label{sec:results}

In this section we present the results of the simulations for the three individual clusters. We study the morphology and use the observed (lack of a) soft excess to constrain the CAB parameters $M$, the inverse coupling, and $\langle E_{CAB}\rangle$, the CAB spectrum mean energy. We then compare the best fit regions for the three clusters to that of the centre of Coma. Throughout we take $H_0=71\text{ km s}^{-1}\text{Mpc}^{-1}$ and $\Omega_{m}=0.27$, to be consistent with \cite{Vacca:2010ss,2012AA...540A..38V}, though we do not expect this choice to have a significant effect on the results compared to that from the systematic uncertainties on the magnetic fields.

\subsection{A665}

No excess is observed from A665, and it is thus an important check for the CAB hypothesis. Since the CAB will always convert to photons in magnetic fields, the conversion probability of an ALP to a photon needs to be small in A665 such that an observable excess is not produced.

A665 will have smaller conversion probabilities than for instance the Coma cluster, where the CAB simulation agreed with the significant soft excess observed, for two reasons. Firstly, the central magnetic field is smaller ($1.3\ \mu\text{G}$ compared to $4.7\ \mu\text{G}$ in Coma), whilst the electron density is roughly similar. Since the conversion probabilities always scale as $P(a\rightarrow\gamma)\propto B^2$ (see Equations \ref{eq:smallangle} and \ref{eq:largedelta}), there is a large $\left(B_{\text{Coma}}/B_{\text{A665}}\right)^2$ reduction in conversion probabilities for A665. Secondly, A665 has larger coherence lengths ($\Lambda_{max,\text{ Coma}} \ll \Lambda_{max,\text{ A665}}$) roughly by a factor 10. With this increase, we see according to Equation \ref{eq:delta} the small $\Delta$ regime will be reached at $\sim 10$ times the impact parameter as for Coma---thus for the energies and impact parameters we are concerned with, the small angle approximation is never reached in A665. Outside of the small angle regime the conversion probabilities scale as 
\beq
P(a\rightarrow \gamma) \propto \frac{1}{L}\left(\frac{B}{n_e}\right)^{2},
\eeq
since the conversion probability per domain is now independent of $L$, but the ALP passes through $D/L$ domains, this should lead to another large reduction in conversion probabilities. However, the conversion in Coma was mostly in the small angle regime, and since probabilities in the small angle formula are naturally smaller (large $\Delta$ implies $\sin^2\Delta=1/2$, whereas small $\Delta$ implies $\sin^2\Delta\approx\Delta^2\ll 1$) the net result is a factor two drop in conversion probabilities for A665 compared to Coma.

From the above two factors the conversion probabilities are a large factor smaller in A665 compared to the previously considered Coma cluster. However, the increased size of A665, both in terms of the cross-sectional area entering the luminosity calculation, and increased propagation distance,\footnote{We note that the finite volume of the simulation naturally affects the results. A larger propagation distance, assuming the magnetic field is still non-zero, will naturally lead to larger conversion probabilities due to the dependence on $D$ of Equation \ref{eq:delta}.} results in a luminosity that is a factor of $\sim 9$ smaller than that for Coma. The morphology is also completely different. In Coma the conversion probabilities were in the small-$\Delta$ regime and were thus larger at low radii, in A665 the opposite is true, see Equation \ref{eq:largedelta}. The CAB conversion is thus largest in the region where the thermal emission from the ICM is smallest.

\begin{figure}
		\centering
\subfloat[]{
		\includegraphics[width=0.49\textwidth]{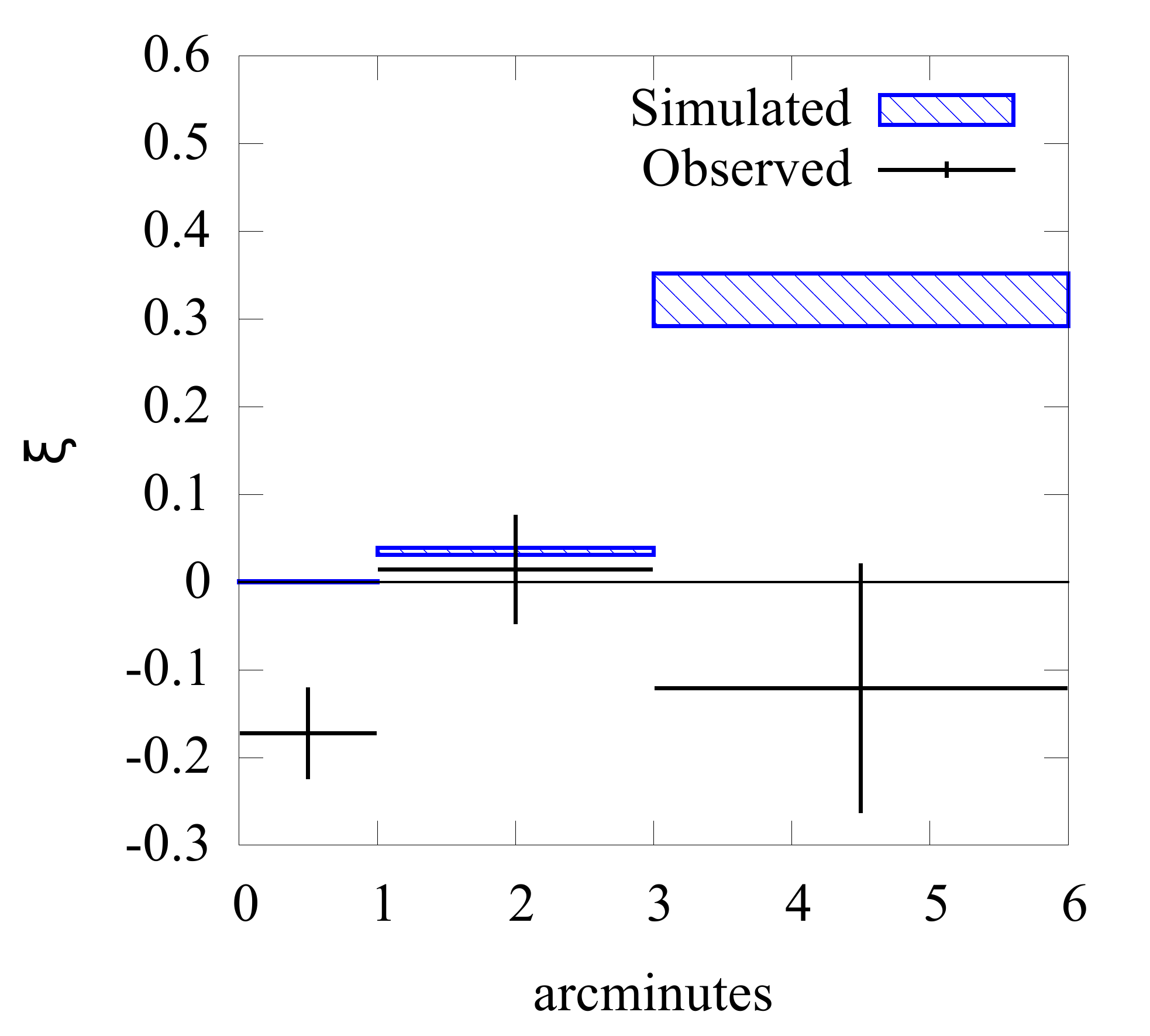}
		\label{fig:a665excess}}
\subfloat[]{
		\includegraphics[width=0.49\textwidth]{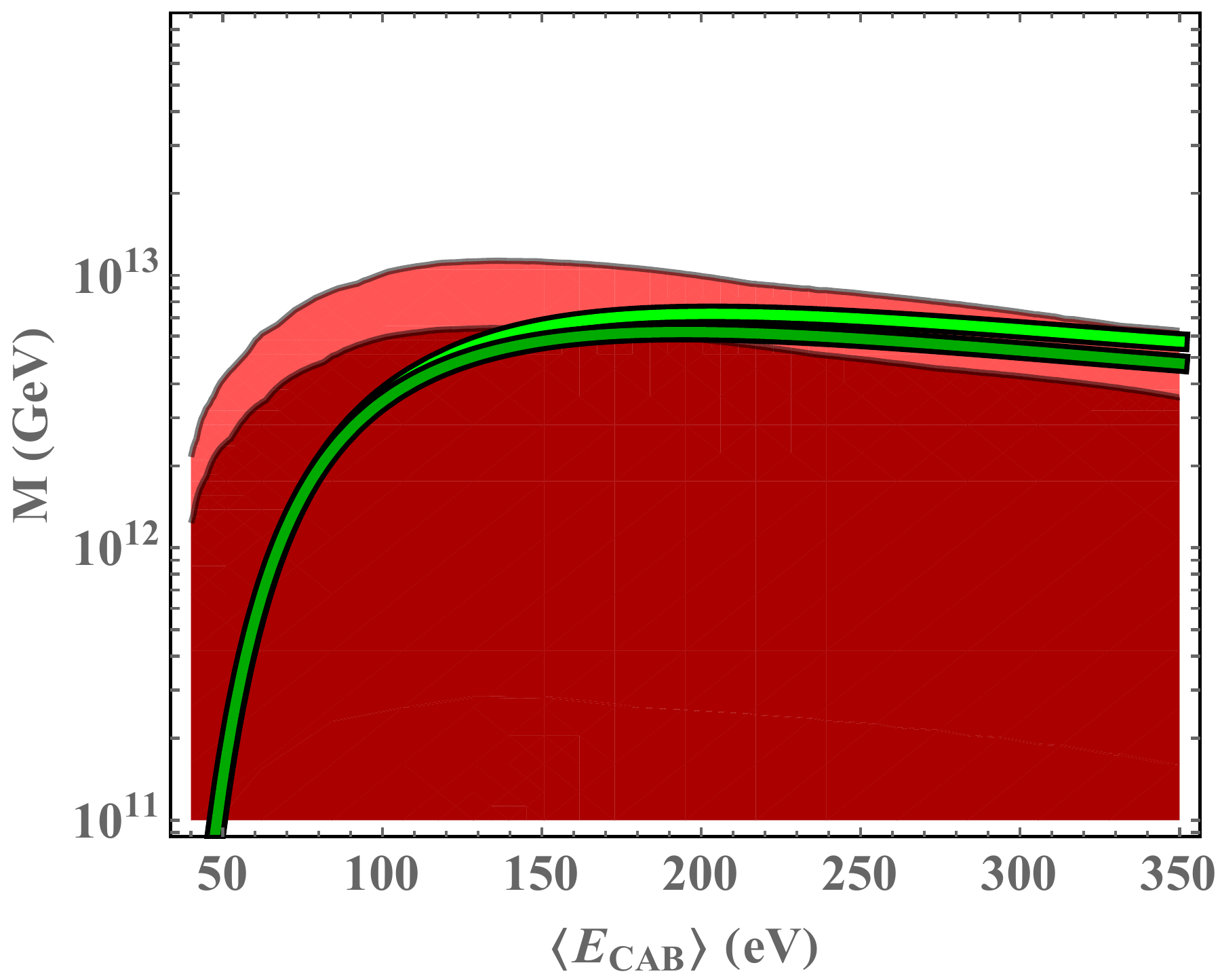}
		\label{fig:a665contours}}
		\caption{a) The simulated and observed fractional excess in A665, the simulation takes $M=7\times 10^{12}\text{ GeV}$ and $\langle E_{CAB}\rangle = 150\text{ eV}$. The observed data error bars are 68\% limits on the observed counts, the shaded region accounts for statistical uncertainty on simulation. b) The $M$ and $E_{CAB}$ parameter space, shown in red is the region where a non-zero soft excess would be observable at 95\% C.L., overlaid in green are the best fit regions from the centre of Coma.}
\label{fig:a665}
	\end{figure}

As the canonical CAB parameter values we set the mean CAB energy to $150\text{ eV}$, and the inverse coupling to $M=7\times 10^{12}\text{ GeV}$. For these values the total fractional excess across the whole cluster is $\xi=0.025$, which is unobservably small. Figure \ref{fig:a665excess} shows the simulated and observed fractional excesses upon dividing the cluster into three annuli---note 6 arcminutes corresponds to 1090 kpc at a distance of A665. The errors on the observed data correspond to 68\% confidence levels on the observed counts. The hot ICM emission at large radii is very small ($9.1\times 10^{-3}\text{ s}^{-1}$) and thus despite the conversion probabilities being suppressed with respect to Coma, the CAB does indeed predict an excess at $>3\text{ arcminutes}$.

The shaded regions for the simulation data points only take into account the statistical uncertainties on the random nature of the field. It was shown in \cite{Angus:2013sua} that upon generating several stochastic magnetic fields with the same power spectra and radial behaviour, the conversion probabilities at most varied by $10\%$. As discussed earlier, there is an additional uncertainty in predicting the counts in the ROSAT detector for a given luminosity. We will take this into account when constraining the CAB parameter space next.

The inverse coupling $M$, and the mean CAB energy $\langle E_{CAB}\rangle$, are unknown parameters in the model. Thus we now redo the analysis scanning over these two parameters, the resultant parameter space is shown in Figure \ref{fig:a665contours}. The red regions signify the parameter space where a significant excess is simulated---i.e. the simulated fractional excess would be distinguishable from zero at 95\% C.L. in \cite{astroph0205473}, this corresponds to a count rate in the outer annulus of $2.6\times 10^{-3}\text{ s}^{-1}$. The light red region corresponds to allowing the count rate to be 50\% larger than PIMMS predicts, the darker region corresponds to allowing the PIMMS prediction to be smaller by 50\%, as discussed in Section \ref{sec:lum}. The shape of the excluded region is easy to understand, for a given value of $M$, the largest luminosity occurs when most of the CAB spectrum is in the R1R2 energy band, hence forcing the tightest bounds on $M$. Note that we cannot push the CAB energy up arbitrarily high in order to not produce an unobserved excess in energy bands above $0.4\text{ keV}$. We see that the red regions include parts (all) of the best fit region found from simulations of the centre of the Coma cluster. The simulation thus predicts a significant soft excess, in the sense defined above, if $M\lesssim 6-10 \times 10^{12}\text{ GeV}$.

The best fit regions from Coma and disfavoured region from A665 thus overlap each other, though we caution that the favoured or disfavoured values of $M$ for both Coma and A665 have an additional uncertainty due to systematics on the magnetic field models.  For instance as discussed earlier these turbulent, stochastic field models are likely to be simplified realisations of the true magneto-hydrodynamical ICM. In addition, the parameters of the model for A665 are constrained only by comparing simulated to observed radio halos, this requires several assumptions about the form of the relativistic electron population within the cluster. As a result, the observed anisotropies and shift in the radio halo centre and X-ray centre are assumed to be caused purely by the stochastic magnetic field (note a field with very large coherence lengths across the cluster will naturally be more anisotropic than one with very small coherence lengths) and not due to the form of the relativistic electrons. These assumptions allow the field parameters to be constrained, however it is likely that the resultant best fit parameters are highly sensitive to the form of the relativistic electron population. The slight overlap of parameter spaces is within the expected uncertainty arising from these concerns.

\subsection{A2199}
	
A2199 is the first cool-core cluster considered in these simulations. The electron density in the centre of the cluster peaks at 0.1 cm$^{-3}$, which is $\sim 30$ times that in Coma, A665 and A2255, and follows a double $\beta$-model profile, given in Equation \ref{eq:doublebeta}. The double $\beta$-model encompasses the extra, cooler (therefore high density) central component of the electron density. We can see from Equations \ref{eq:convprob} and \ref{eq:theta} that outside of the small-$\Delta$ approximation \beq
P(a\rightarrow\gamma)\propto n_e^{-2},
\eeq
and thus high electron densities suppress conversion. The cluster however has a very small core radius ($\mathcal{O}(20\text{ kpc})$ compared to $\mathcal{O}(300\text{ kpc})$ for Coma, A665 and A2255), and thus the electron density drops very rapidly with increasing radius, and thus we do not expect much suppression of conversion probabilities with respect to non-cool-core clusters. As we have argued before, large electron densities imply large magnetic fields, and thus A2199 has a large central magnetic field value of $B_0\sim 12\ \mu\text{G}$, much larger than the two other cluster's magnetic fields. However again this will drop rapidly from the centre due to the small core radius.

\begin{figure}
		\centering
\subfloat[]{
		\includegraphics[width=0.48\textwidth]{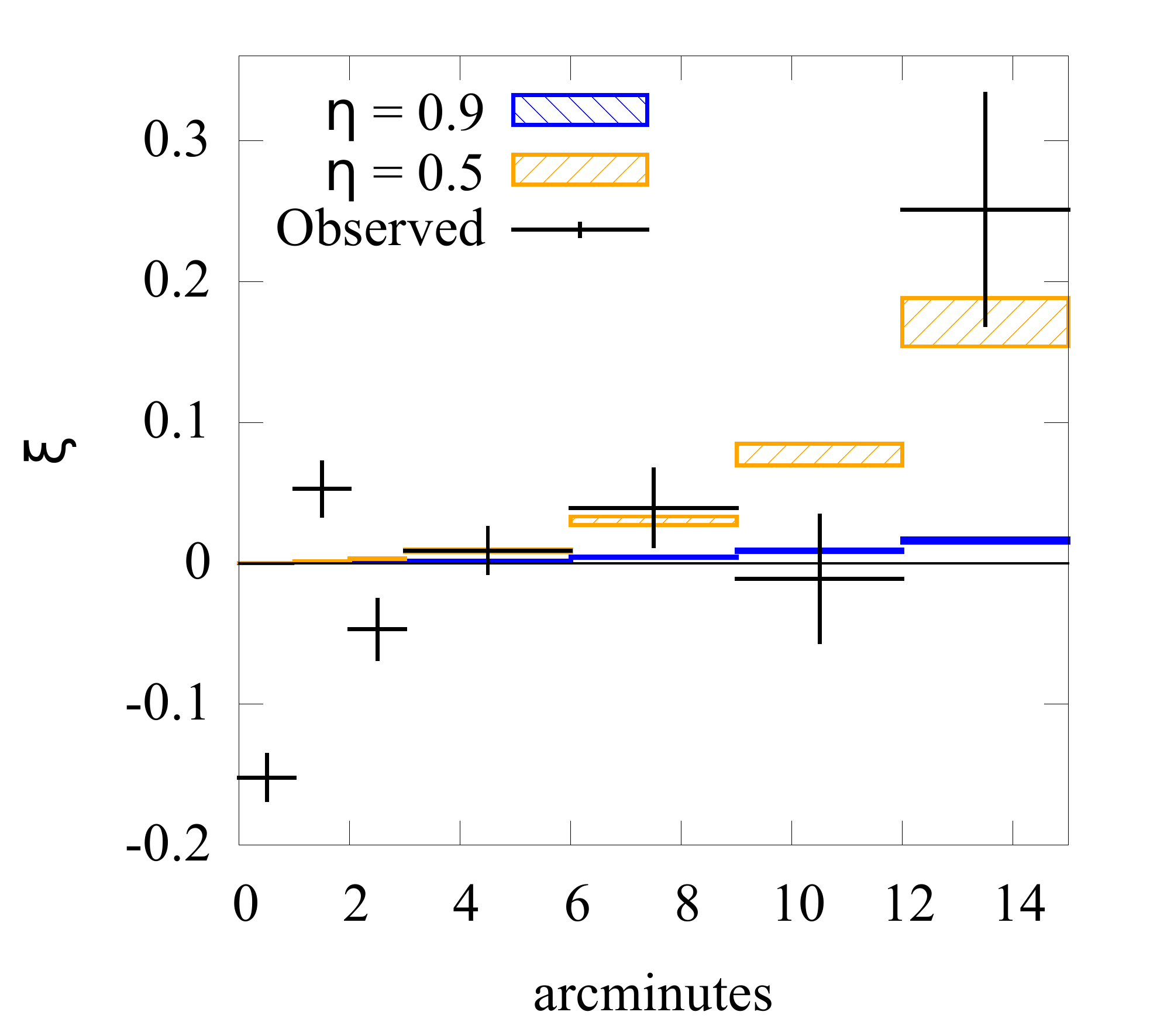}
\label{fig:a2199frac}}
\subfloat[]{
	\includegraphics[width=0.48\textwidth]{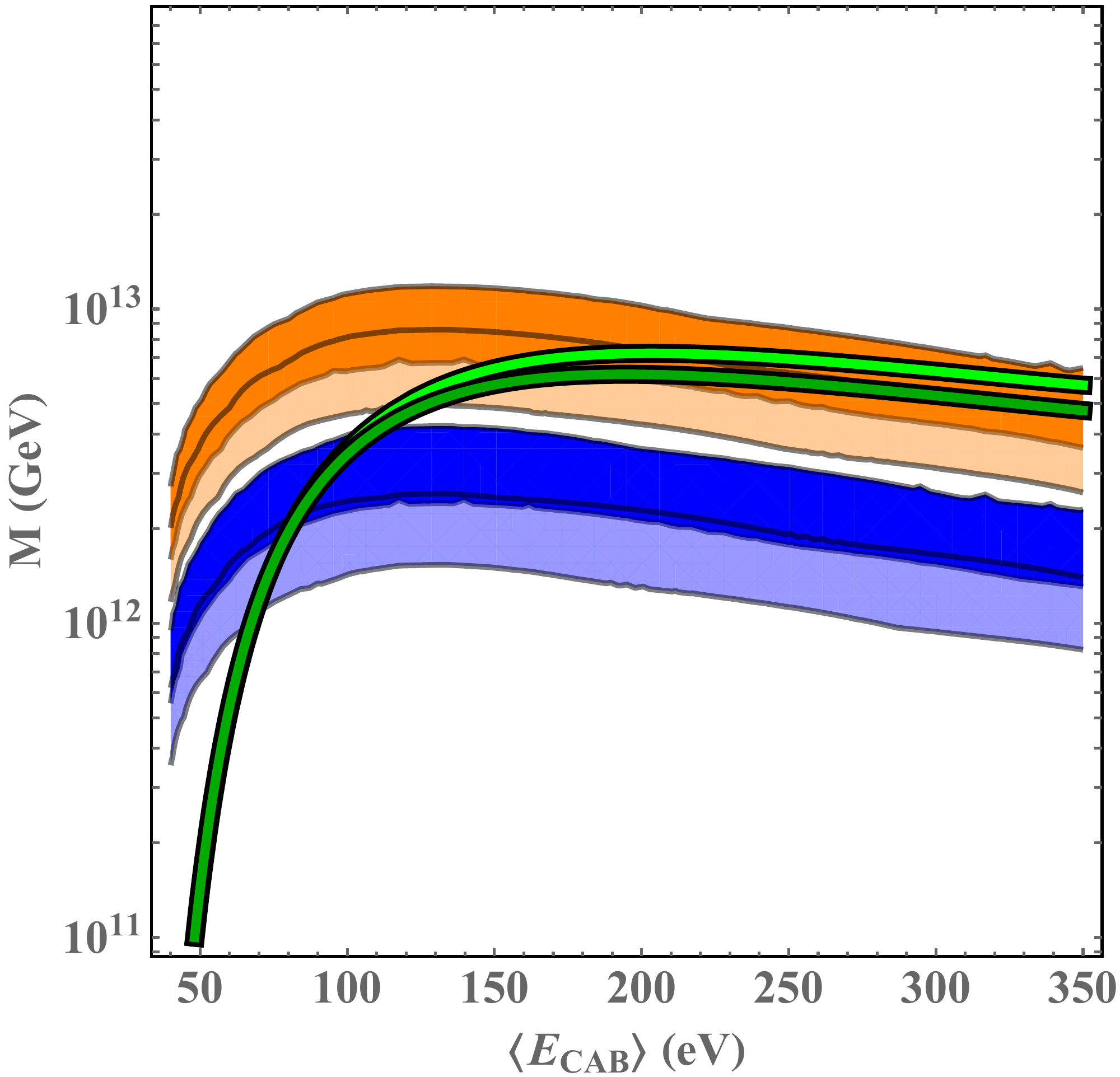}
\label{fig:a2199contours}}
		\caption{a) The fractional excess in A2199. Both simulated data sets are taken with $M=7\times 10^{12}\text{ GeV}$ and $\langle E_{CAB}\rangle = 150\text{ eV}$. b) The $M$ v $\langle E_{CAB} \rangle$ parameter space which fits the data for the two models, defined as when the total observed fractional excess is equal to the total simulated fractional excess between $1-15$ arcminutes. The lighter colours indicate the regions where the excess is fixed to that in the $12-15\text{ arcminutes}$ annulus. For comparison, overlaid in green are the best fit regions from the centre of Coma.}
		\label{fig:a2199}
\end{figure}

The simulated fractional excess in A2199 for the canonical magnetic field parameter values $(B_0,\ \eta)=(11.9\ \mu\text{G},\ 0.9)$ can be seen as the blue points in Figure \ref{fig:a2199frac}---15 arcminutes at A2199 corresponds to $\sim 540\text{ kpc}$. In this plot $M=7\times 10^{12}\text{ GeV}$ and $\langle E_{CAB}\rangle =150\text{ eV}$ as usual, and again the shaded region takes into account the statistical uncertainty on the stochastic field. The CAB-generated luminosity in the $0.2-0.4\text{ keV}$ band between $12-15\text{ arcminutes}$ is simulated to be $4.7\times 10^{39}\text{ erg s}^{-1}$, which is more than an order of magnitude smaller than the observed luminosity (assuming thermal emission) of the soft excess, in the same energy range, of $1.7\times 10^{41}\text{ erg s}^{-1}$ \cite{astroph0205473}. Thus for the Coma best fit parameters the CAB predicts no observable excess in A2199 for the canonical magnetic field parameters.

We first discuss changes to the magnetic field parameters before scanning over the CAB parameter space. As we mentioned earlier, the magnetic field in A2199 is poorly constrained. To be specific, the central magnetic field magnitude and the radial parameter $\eta$ are constrained to be $B_0=(11.9\pm 9.0)\mug$ and $\eta= 0.9\pm 0.5$. In the following we will consider varying these two parameters. These two parameters are degenerate with respect to Faraday rotation measurements, a larger central magnetic field value with a steeper fall with radius (higher $\eta$) produces a magnetic field with the same average value. Thus these two parameters must be changed in tandem. We do not consider any changes in the spectral index or the largest lengthscale $\Lambda_{max}$, although we note that these also have a range of allowed values. 

Within the uncertainties on $(B_0,\ \eta)$ we also show the simulated soft excess for the parameter choice $(B_0,\ \eta)=(6\ \mu\text{G},\ 0.5)$ plotted as the orange points in Figure \ref{fig:a2199frac}. We choose the value $\eta=0.5$ because as mentioned earlier, this corresponds to the case where the energy density in the magnetic field is proportional to the energy density of the thermal gas. Lowering the parameter $\eta$, reduces the rate at which the magnetic field magnitude falls with radius and thus increases the magnetic field magnitude at large radii, thereby increasing the conversion probabilities at large radii. We see this as an increased luminosity and hence increased soft excess at large radii. For these magnetic field parameters we see that the CAB parameters that fit Coma match the observed morphology (given the amount of scatter in the data) and magnitude of the soft excess in A2199 well.

We next fit the magnitude of the soft excess by scanning over the $\left(M,E_{CAB}\right)$ parameter space. We neglect the data in the $0-1\text{ arcminutes}$ annulus due to the presence of a large deficit of soft X-rays. We show the region of CAB parameter space where the simulation fits the data, for both magnetic field parameters mentioned above, in Figure \ref{fig:a2199contours}, where the thickness of the bands takes into account the count rate uncertainty discussed earlier. The dark regions are the regions where the total excess is reproduced, shown in lighter colours are the regions where the magnitude of the excess in the $12-15\text{ arcminutes}$ annulus only is reproduced. We see then that given the uncertainty on the magnetic field strength and radial profile, the excess can be fit for a large range in the inverse coupling, $M$. The CAB converting to photons in A2199 is consistent with Coma and A665 if the magnetic field magnitude falls less steep with radius then the best fit field.

The current soft X-ray data contains a large amount of scatter and uncertainty. The total excess over the whole of the cluster is small (several \%) and not significant, however there are radial regions where there is an observed excess, and there is a general trend that it increases with radius. We have shown that if the magnetic field falls less steeply with radius, the magnitude and general morphology are fit well, for a CAB with $M=6-12\times 10^{12}\text{ GeV}$. However there is large uncertainty in both the magnetic field and the soft excess observations. Clearly a more constrained magnetic field model, and hopefully future soft X-ray observations will shed more light on the CAB in A2199.

\subsection{A2255}
	
The simulated magnetic field in A2255 is an unrealistic `stitching' together of two fields with different coherence lengths. For radii less than the core radius, the field has a flat power spectrum with the full range of scales available. For radii greater than the core radius, the field has a power spectrum index of $\zeta=2$ and a minimum lengthscale of $64\text{ kpc}$. A larger spectral index puts more of the power in smaller momentum modes. As a result the field in the centre of the cluster will on average have much shorter coherence lengths than in the outer parts of the cluster, which will typically have very large coherence lengths. We use a smoothing function at the core radius to interpolate between the two fields.

This magnetic field model then gives a unique morphology to the ALP$-$photon conversion probabilities. For ALPs whose impact parameter is small, $r<r_c$, the ALPs will pass through a large region where the field has small coherence lengths. The ALP conversion is then in the small-angle regime for a large proportion of the ALP's propagation, thus we expect conversion probabilities that decrease with increasing impact parameter, until $r\sim r_c$. For impact parameters larger than the core radius the ALPs will propagate only through the $\zeta=2$ magnetic field, and the large coherence lengths here result in conversion probabilities which are out of the small $\Delta$ regime (although high energies and large impact parameter will reduce the angle $\Delta$ back into the small angle regime), and thus the conversion probabilities increase with radius.

\begin{figure}[t]
		\centering
		\includegraphics[width=0.65\textwidth]{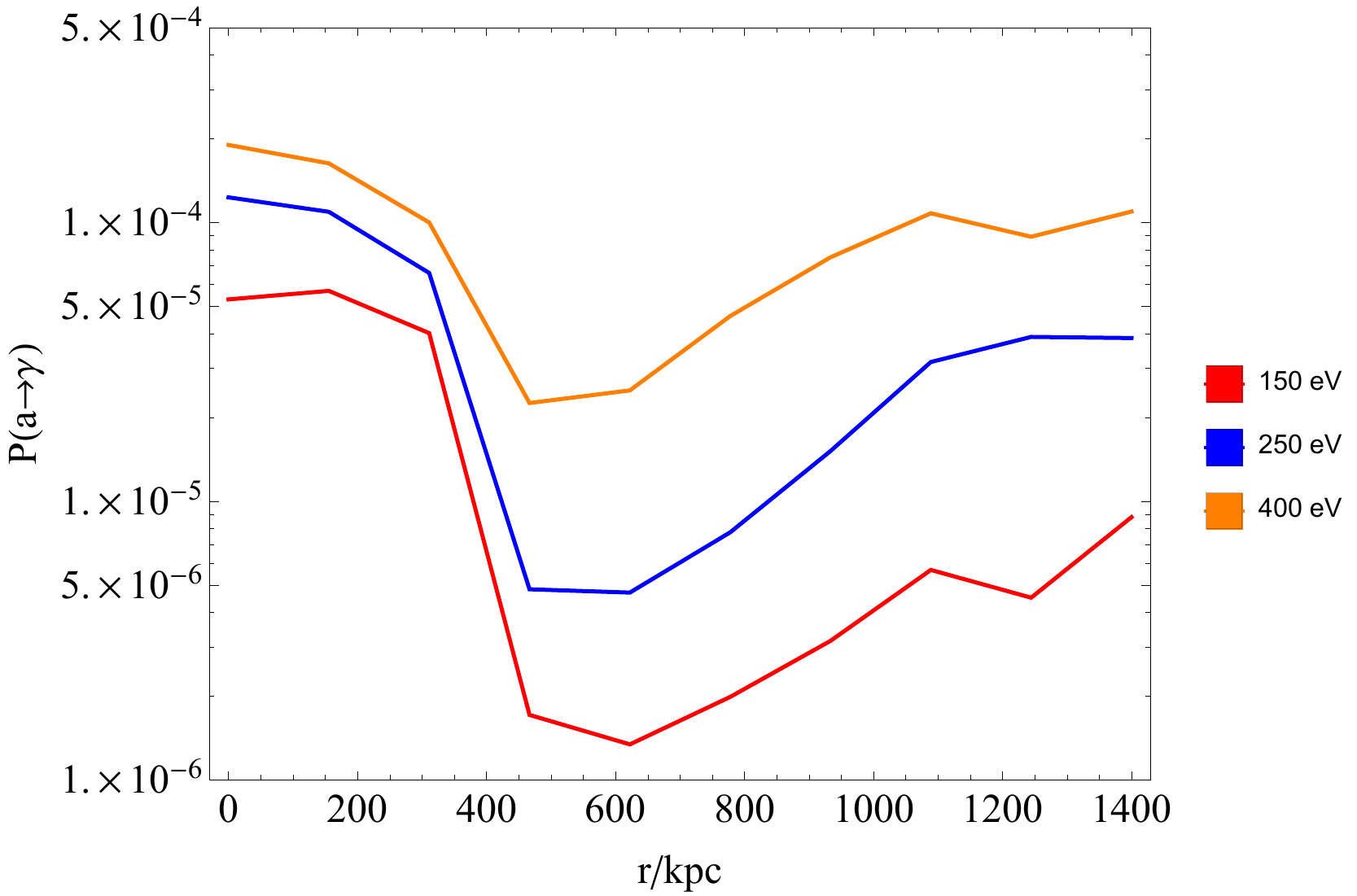}
		\caption{The ALP-photon conversion probabilities in the cluster A2255 for an ALP of impact parameter $r$, shown for three different ALP energies. We see the change in behaviour between ALPs that propagate mostly through the flat spectrum field at low impact parameter, and those that propagate only through the $\zeta=2$ field at large impact parameter. The conversion probabilities are computed with $M=7\times 10^{12}\text{ GeV}$.}
		\label{fig:a2255probs}
\end{figure}

We show this feature of the conversion probabilities for three ALP energies in Figure \ref{fig:a2255probs}. We see the clear feature at $r_c\sim 400\text{ kpc}$ where the ALPs are no longer passing through any of the smaller coherence length field. The conversion probabilities in the outer field are smaller relative to the centre due to the large coherence lengths resulting in the ALPs passing through few magnetic field domains. 

\begin{figure}
	\centering
	\subfloat[]{
		\includegraphics[width=0.48\textwidth]{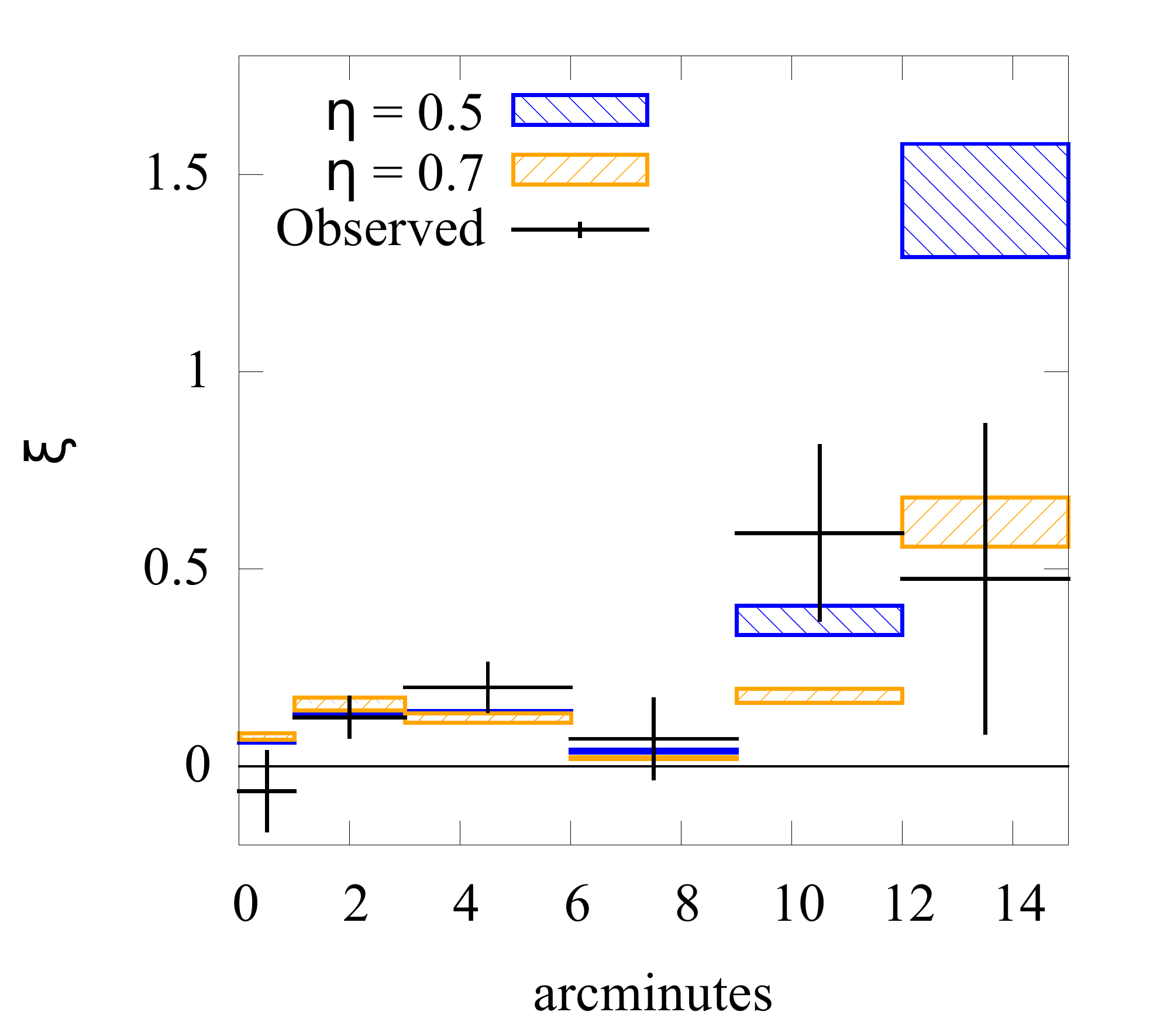}
\label{fig:a2255frac}}
	\subfloat[]{
		\includegraphics[width=0.48\textwidth]{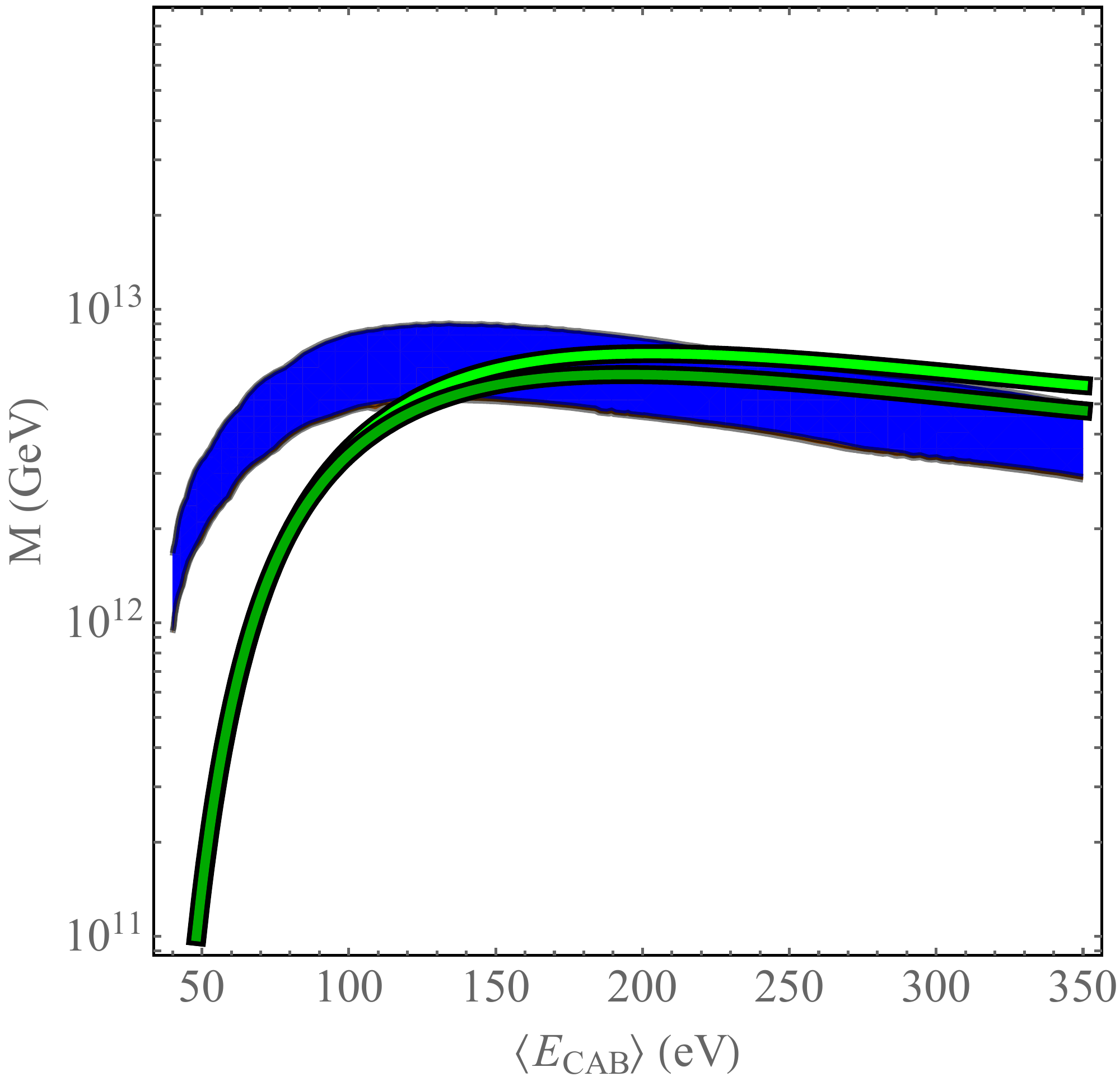}
\label{fig:a2255contours}}
		\caption{a) The simulated and observed fractional excess in A2255. Plotted are the magnetic field models with radial parameter $\eta=0.5$ in blue, and $\eta=0.7$ in orange. For the simulations we take $M=7\times 10^{12}\text{ GeV}$ and $\langle E_{CAB}\rangle = 150\text{ eV}$. b) The $M$ v $\langle E_{CAB} \rangle$ parameter space which fits the data in the central radial zones (excluding $>9$ arcminutes where the data is uncertain) for the two models (which overlap almost completely). For comparison, overlaid in green are the best fit regions from the centre of Coma.}
\label{fig:a2255}
\end{figure}

In Figure \ref{fig:a2255frac} we plot the simulated soft excess for the two radial parameters $\eta=0.5,\,0.7$.\footnote{Note the magnetic field magnitude normalisation in A2255 is chosen slightly different to that of A2199. According to \cite{2006AA...460..425G} we normalise the field such that the average magnetic field inside the central $1\text{ Mpc}^3$ is equal to $1.2\ \mu\text{G}$, which automatically ensures the value of $B_0$ changes according to the value of $\eta$ chosen.} Since the value of the radial parameter $\eta$ is not well constrained we choose the these two values as representative of the two interesting cases discussed earlier. In the plot we have again taken $M=7\times 10^{12}\text{ GeV}$ and $\langle E_{CAB}\rangle=150\text{ eV}$, and the shaded region is the statistical uncertainty. 15 arcminutes at A2255 corresponds to $\sim 1.4\text{ Mpc}$. We see that both field parameter choices reproduce the soft excess magnitude, and the morphology at radii $<9\text{ arcminutes}$. The low conversion probabilities around $\sim 500\text{ kpc}$ translate into a small, unobservable excess, exactly where the observations reveal no excess between $6-9\text{ arcminutes}$.

Both of the models then have large excesses at large radii due to the radial profile of the conversion probabilities and the low expected ICM emission. The data at $>9\text{ arcminutes}$ are uncertain and thus the morphology is not a concern, however the magnitude of the simulated excess for $\eta=0.5$ at the outskirts is much higher than that observed. Increasing the radial parameter $\eta$ results in the magnetic field dropping off more rapidly with radius, lower field values necessarily produce lower conversion probabilities and thus increasing to $\eta=0.7$ does not overproduce soft X-rays and fits the data at $>9$ arcminutes well.

In Figure \ref{fig:a2255contours} we illustrate the fact that the CAB parameters that best fit the observed soft excess between $1-9$ arcminutes overlap the best fit CAB parameter region from the central part of Coma. Again the thick band takes into account varying the count rate prediction from PIMMS by $\pm 50\%$. We do not fit the large radius data due to the signal uncertainties mentioned earlier. The CAB morphology of the soft excess in A2255 clearly prefers a field which falls with radius more steeply than the canonical $B^2\propto n_e$ choice.

Finally we note that the simulated magnetic field is not a realistic field. In addition to the usual systematic uncertainty due to assuming this stochastic model for the cluster magnetic field, the simulated field for A2255 is produced by stitching together two stochastic fields produced with different power spectra. Thus while the morphology of the simulated soft excess is in very good agreement with the observed excess, it is clear that this is mostly a feature of the field model used. The magnetic field transforms from a field with small coherence lengths to one with large coherence lengths over a short distance, and thus the conversion probabilities drop suddenly beyond the core radius. A better implementation of the field would be to have increasing coherence lengths as a function of radius, something that currently cannot be implemented as part of the simulation.  A magnetic field with a spectral index which increases with radius and thus has coherence lengths which smoothly increase with radius would not have such striking distinction between small and large impact parameter as in Figure \ref{fig:a2255probs}. It is likely that the CAB produced excess between $6-12\text{ arcminutes}$ would not be as small, but would still provide a good fit to the observed morphology. In \cite{Kraljic:2014yta} the outskirts of Coma were analysed using a magnetic field whose coherence lengths increase with radius using a semi-analytical approach, we expect such a method is better suited to studying A2255.

\subsection{Comparison and Summary}

Here we compare the CAB parameter space between the clusters, and summarise our results. Firstly we note the dependence of the results on $\Delta N_{\text{eff}}$. The luminosity of the soft excess produced by a CAB is proportional both to the flux of ALPs passing through the cluster per second, and the conversion probability. Since the CAB energy density is proportional to $\Delta N_{\text{eff}}$ and the conversion probabilities are always proportional to $M^{-2}$, the soft X-ray luminosity from a CAB behaves as $\mathcal{L}\propto \frac{\Delta N_{\text{eff}}}{M^2}$.

Observations of the cluster A665 have shown no excess, thus we can use this to bound the CAB parameter space. The simulation predicts a low total excess across the cluster, consistent with zero. However, a large excess is predicted at large radii, since here the conversion probabilities are highest, and the ICM emission is lowest. Stipulating that the simulated excess from a CAB at large radii should not be larger than the expected emission at 95\% confidence, results in the disfavoured region
\beq
M<6-10\times 10^{12}\text{ GeV},
\eeq
for CAB mean energies $\langle E_{CAB}\rangle\sim 100-250\text{ eV}$. The range takes into account the uncertainty on where the bound lies due to the uncertainty in predicting count rates in the ROSAT PSPC detector given the simulated luminosity. This disfavoured region overlaps with the preferred region from the centre of Coma ($M=5-8\times 10^{12}\text{ GeV}$), and thus the simulation of a CAB in A665 is in slight disagreement with the simulations of the Coma cluster. However the systematic uncertainties on the magnetic field models results in the actual bounds being uncertain  to at least the level of this overlap, and thus this small overlap is acceptable.

For the cluster A2199, a CAB produces a soft X-ray excess which would be very small (a factor of $\sim 20$ smaller than observed) for the canonical radial scaling parameter value of $\eta=0.9$. We can alleviate this issue by decreasing this parameter within its allowed uncertainty, to $\eta=0.5$. This results in a magnetic field magnitude which falls with radius more slowly and thus has a much more rapidly increasing soft excess with radius. The choice $\eta=0.5$ then gives a good fit to the morphology of the observed excess. The magnitude of the excess is fit well for the lower $\eta$ field for the parameters $M=6-12\times 10^{12}\text{ GeV}$ and thus agrees with the analysis of Coma.

The simulations of A2255 also fit the observed excess well. The canonical choice of radial parameter $\eta=0.5$, overproduces X-rays in the outer regions, but changing this value to $\eta=0.7$ (which is allowed within observational uncertainties) produces an excess with the right morphology and magnitude for $M=5-9\times 10^{12}\text{ GeV}$, which agrees with A2199 and again with the Coma cluster. The morphology is fit very well, but this is mostly a feature of the magnetic field model used. The magnetic field model has an unrealistic sharp transition around the core radius from small coherence lengths to large coherence lengths. Where this transition happens the excess is small, exactly where the observed excess is small.

\begin{figure}[t]
	\centering
\subfloat[]{
	\includegraphics[width=0.49\textwidth]{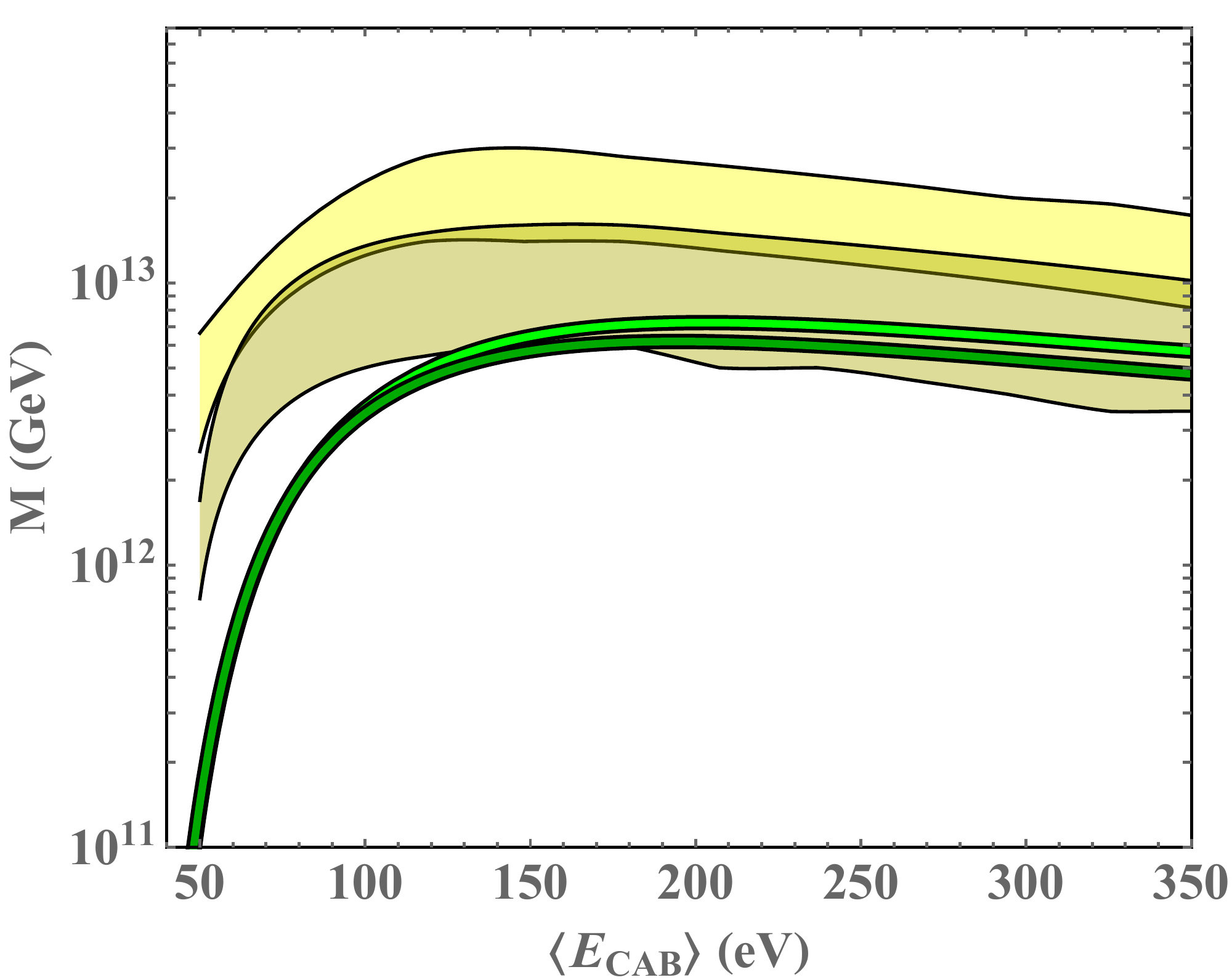}
\label{fig:coma}}
\subfloat[]{
	\includegraphics[width=0.49\textwidth]{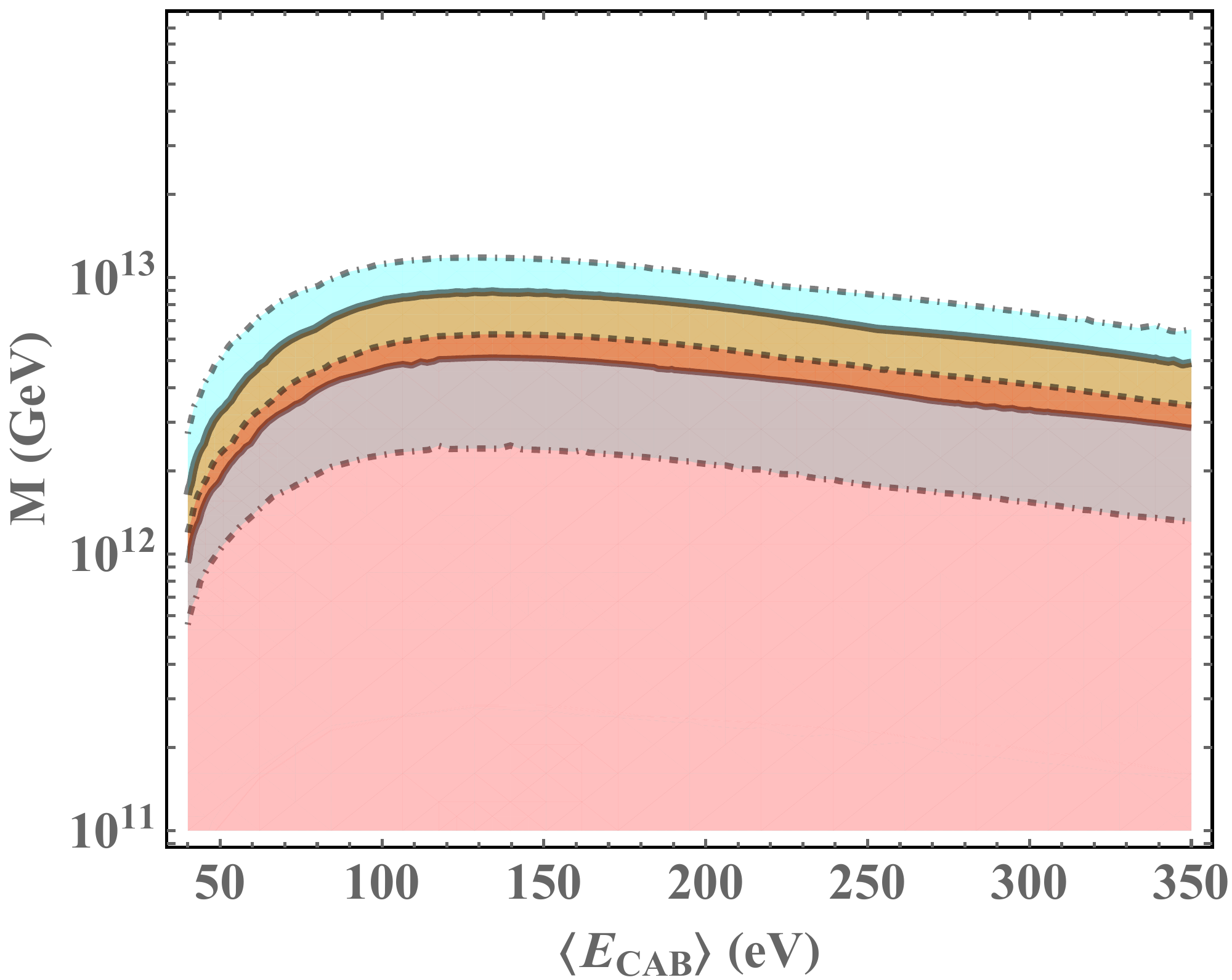}}
\\
\subfloat[]{
	\includegraphics[width=0.55\textwidth]{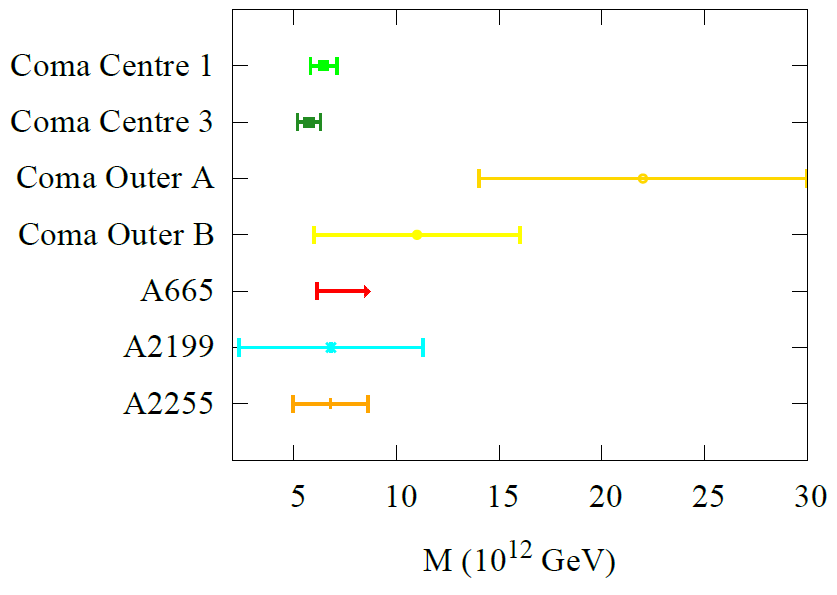}
\label{fig:concordance}}
\caption{a) The best fit regions of the $M$ vs $\langle E_{CAB}\rangle$ parameter space from the analyses of the centre (green) and the outskirts (yellow) of the Coma cluster. The light green and dark green regions correspond to Models 1 and 3, of the centre, respectively. Similarly the light yellow and dark yellow regions are Model A and Model B from the outskirts of Coma (which correspond to Models 1 and 3 from the centre). b) The best fit regions from the three clusters studied here: the disfavoured region from A665 is shown in red (dashed line); the best fit region for A2199 is shown in blue (dot-dashed line); and the same for A2255 is in orange (solid line). c) A comparison of the best fit values for $M$ between simulations of the three clusters studied here, along with the best fit values from the Coma studies. The values of $M$ have been taken corresponding to (the arbitrary choice) $\langle E_{CAB}\rangle =150\text{ eV}$.}
\label{fig:allcontours}
\end{figure}

In Figure \ref{fig:allcontours} we show the regions of the CAB parameters which fit the soft X-ray observations of the three clusters considered here, for comparison we also show the CAB parameter space which fits the soft excess in the centre and the outskirts of the Coma cluster. We show the good agreement between clusters by showing in Figure \ref{fig:concordance} the best fit values of $M$ for all the clusters considered so far, setting (arbitrarily) $\langle E_{CAB}\rangle=150\text{ eV}$. From the simulations of these three clusters we find the preferred range of the coupling $M$ to be,
\beq
M=6-12\times 10^{12}\text{ GeV}\ \sqrt{\frac{\Delta N_{\text{eff}}}{0.5}}.
\eeq
Which overlaps well with the parameter space from the centre of Coma, $M= 5-8\times 10^{12}\text{ GeV}$, and from the outskirts of the Coma cluster, which requires $M=5-30\times 10^{12}\text{ GeV}$.

It has been observed that the cluster soft excess is preferentially seen at large radii \cite{astroph0205473}. In all three clusters we have studied in this paper, the fractional excess produced from CAB conversion is largest at large radii. This is because the thermal emission from the ICM falls of faster than the conversion probabilities from the magnetic field. Thus we see that the CAB re-produces this general morphological trend.

\section{Conclusions}
\label{sec:conclusions}

In this paper we have continued the analysis of whether a primordially-generated background of relativistic ALPs of energy $\sim 0.1-0.4\text{ keV}$ can explain an excess of soft X-rays seen from many galaxy clusters. This work is an extension of \cite{Angus:2013sua,Kraljic:2014yta} which have shown that the soft X-ray excess in the Coma cluster, out to $5\text{ Mpc}$, can be explained by such a cosmic ALP background (CAB) converting to photons in the cluster's magnetic field. Here we have analysed the CAB hypothesis in A655 which shows no excess emission, to check whether the CAB produces too large a flux of soft X-rays. We have also analysed the magnitude and morphology of the soft excesses in A2199 and A2255.

We calculated the expected luminosities from CAB conversion by modelling the cluster magnetic field as a tangled, stochastic field with power-law power spectrum and magnitude which falls as a power of the electron density of the ICM, and numerically calculating the conversion probability for an ALP of mass $m_a<10^{-13}\text{ eV}$ travelling through the field.  We compared the simulation results to ROSAT PSPC count rates in the $0.1-0.28\text{ keV}$ R1R2 band.

Since the magnitude of the soft X-ray excess in the clusters is dependent on the CAB parameters $M$, the inverse coupling, and $\langle E_{CAB}\rangle$ the CAB spectrum mean energy, we are able to use the simulations to constrain the CAB parameter space. We have confirmed that the three clusters studied here agree on the CAB parameter space, and agree with that obtained from the analyses of Coma. The favoured value of the inverse coupling $M$ from this analysis is
\beq
6\times 10^{12}\text{ GeV}<M<1.2\times 10^{13}\text{ GeV},
\eeq
for $\langle E_{CAB}\rangle \approx 100-250\text{ eV}$, which is consistent with the analysis of the Coma cluster. The cluster A665 is in slight tension with the other clusters, since an excess is produced at large radii for values $M< 6-10\times10^{12}\text{ GeV}$, which takes into account the uncertainty in computing the expected count rate in the ROSAT R1R2 band for given a soft X-ray source luminosity. There is additional uncertainty due to systematic uncertainties on the cluster magnetic fields thus this small disagreement is acceptable. The morphology of the soft excesses in A2199 and A2255 are both reproduced well by CAB conversion to photons.

While the simulations conducted so far have shown good agreement between CAB parameters across the clusters studied, the systematic uncertainties in the magnetic fields of clusters limit further analysis. The four clusters studied so far exhaust the list of clusters for which both soft X-ray excess observations and detailed magnetic field power spectra and radial profiles are constrained. On top of this it is likely that these magnetic fields models do not capture the full magneto-hydrodynamical structure of the real cluster magnetic fields. The magnetic field models used represent the most detailed models of cluster magnetic fields which have been compared to data, however they are still limited due to the limited knowledge attainable from Faraday rotation and the assumptions needing to be made about the relativistic electron population for analysis of the magnetic field using radio structures. Since the CAB gives a testable prediction given knowledge of the cluster magnetic field, better knowledge of cluster magnetic fields will allow us to test the CAB hypothesis further. 

\section*{Acknowledgements}
I thank Pedro Alvarez, Stephen Angus, Joseph Conlon, David Kralji\v{c}, David Marsh, Markus Rummel and Lukas Witkowski for helpful discussions regarding this work. I thank David Kralji\v{c} and Markus Rummel for the data for Figure \ref{fig:coma}. I would also like to thank Joseph Conlon for carefully reading and commenting on the manuscript. I am funded by an STFC studentship.

\bibliographystyle{JHEP}
\bibliography{CAB}

\end{document}